\documentclass{article}

\pdfoutput=1

\usepackage[utf8]{inputenc}
\usepackage[square,sort,comma,numbers]{natbib}
\usepackage{graphicx}
\usepackage{xspace}
\usepackage{amssymb,amsmath}
\usepackage{subcaption}
\usepackage{authblk} 

\newcommand{\LaO}{La\subt{2}O\subt{3}\xspace}
\newcommand{\LaB}{LaB\subt{6}\xspace}

\newcommand{\supt}[1]{\textsuperscript{#1}}
\newcommand{\subt}[1]{\textsubscript{#1}}
\newcommand{\BfC}{B\subt{4}C\xspace}
\newcommand{\Heplus}{He\supt{+}\xspace}
\newcommand{\degree}{$^{\circ}$\xspace}

\title{Limits of surface analysis of thin film compounds using LEIS}

\author[1]{Andrey A. Zameshin}
\author[1]{Andrey E. Yakshin}
\author[1]{Jacobus M. Sturm}
\author[1]{Cristiane Stilhano Vilas Boas}
\author[1]{Fred Bijkerk}

\affil[1]{Industrial Focus XUV Optics Group, MESA+ Institute for Nanotechnology, University of Twente, Drienerlolaan 5, 7522 NB Enschede, The Netherlands}

\date{06 September 2018}

\begin{document}

\maketitle

\begin{abstract} 

Low Energy Ion Scattering (LEIS) was employed to study the surface composition of thin films of Ru on B, C and B4C films at different stages of growth. Effects of surface segregation of C were observed. Previously unknown matrix effects were observed in these samples, expressed in the decrease of LEIS signals of Ru, B and C at low Ru concentrations. The effect disappears for Ru-rich surfaces. Measurements with different He+ ion energies prove that the characteristic velocities of the elements involved change with surface composition. We suggest that these matrix effects appear due to the changes in neutralization efficiency in quasiresonant neutralization from the valence band (VB-qRN). This neutralization channel is present in elemental C and B due to a wide valence band with energy states as low as -20 eV, which are in a (quasi-)resonance with the He 1s level. This mechanism was earlier reported for graphitic carbon. We suggest that it can be applied to a much wider range of materials, leading to potential matrix effects in LEIS from a variety of surfaces, containing B, C and potentially O and N atoms, e.g. borides, carbides, oxides and nitrides, as well as alloys with B and C. This hypothesis is supported by additional LEIS measurements on oxidized Ru which show matrix effect in Ru-O LEIS signals as well. We argue that it is possible to avoid the matrix effects from compounded surfaces within certain ranges of composition by a proper choice of reference samples, while for other compositions knowledge of characteristic velocities is required for reliable quantification.

\end{abstract}

\section{Introduction} \label{RLsec:Introduction}

Low Energy Ion Scattering (LEIS) is a surface analysis technique with extremely low information depth, normally only the topmost atomic layer~\cite{Brongersma2007, Cortenraad2001crystalface, Primetzhofer2011}. Extreme surface sensitivity is given by very efficient neutralization processes of noble gas ions with energies below 8 keV. The downside of this is the dependence of the LEIS signal on neutralization efficiency. Usually the neutralization efficiency of \Heplus from a given surface atom does not depend on the surrounding atoms, in which case quantification of surface composition with LEIS is possible by comparing LEIS signals between given a sample and a reference sample with known surface composition and density~\cite{Brongersma2007}. However, occasionally neutralization processes and therefore LEIS signals can depend on the surrounding atoms. This phenomenon is called a matrix effect, and it can severely complicate the surface quantification. It is often considered that cases of matrix effects are few and limited to very specific situations.

In an earlier paper LEIS was used for studying the growth process of magnetron sputtered thin films of Ru on amorphous Si~\cite{ColomaRibera2016}. In this paper we extend this research to growth of thin films of Ru on several other amorphous substrates, namely B, C and \BfC. The application of Ru thin films include catalysis~\cite{Saadatjou2015}, seeding and substrate layers for Cu~\cite{Yeo2013, Hsieh2015} or graphene~\cite{Sutter2008, Wintterlin2009}, bottom and top electrodes for DRAM capacitors~\cite{Kim2004}. Growth of Ru thin films on B, C and \BfC substrates is important for a variety of multilayer structures for reflection of soft and hard X-Rays~\cite{Stearns1991,Spiller1994proceedings, Ksenzov2010}. Interface width and composition of such systems is important to their performance, therefore they can benefit from the development of new techniques of thin film interface analysis.

Analogous to the LEIS studies of the Ru/Si combination~\cite{ColomaRibera2016}, the thin film deposition and LEIS measurements were performed without breaking vacuum. This allows to perform almost \textit{in-situ} quantification of the surface composition for different thicknesses of Ru, and therefore compose a deposition depth profile (growth profile)~\cite{ColomaRibera2016}, as opposed to the more conventional sputter depth profile, which is produced by removing the top layer by sputtering and is consequently affected by sputter-induced intermixing and preferential sputtering. 

An important aspect of this paper is related to the discovery of new matrix effects in Ru-B and Ru-C material combinations, and the limitations that these matrix effects impose on the quantification of Ru-B and Ru-C based thin films. We also study the origin of these matrix effects on the example of the Ru-B combination, and further extrapolate the findings to a much wider range of compounds. The results of this work are separated into two parts. In the first part (Section~\ref{RLsec:LEIS_3keV}), LEIS measurements with 3 keV \Heplus for different Ru thicknesses are used to obtain deposition depth profiles, but it becomes evident that in a thickness range below 1 nm the quantification procedure cannot be trusted. In the second part of the paper (Section~\ref{RLsec:varE_measurements}), the suspected matrix effect is studied for the Ru-B system by measuring LEIS signals at different energies.

\section{Experimental} \label{RLsec:Experimental}

All samples studied in this work have been produced in a home-designed UHV magnetron sputtering chamber with a base pressure up to $ 2 \times 10^{-10}$ mbar and target-to-substrate distance of 8.6 cm. Depositions were performed with $ 1 \times 10^{-3}$ mbar of Kr as a sputter gas. DC magnetron sputtering was used for Ru, \BfC and C, while RF magnetron sputtering was chosen for B and Si. The thicknesses of the deposited thin films were controlled by quartz crystal microbalances (QCM). QCM readings were calibrated in advance, using a set of calibration thin films of which the thicknesses were extracted from fitting of CuK$\alpha$ Grazing Incidence X-Ray Reflectivity curves. 

The deposition chamber was connected with a UHV transfer system to the LEIS chamber, sample transfer taking about 10 minutes between the deposition and the measurement. The LEIS setup was a Qtac100 instrument by ION-TOF, with a base pressure down to $ 1 \times 10^{-10}$ mbar, equipped with a double toroidal electrostatic analyzer and an electron impact ion source, with ion incidence angle normal to the surface of a sample and a scattering angle of 145\degree. All LEIS spectra are obtained with \Heplus primary ions. The default \Heplus energy was 3 keV, and in a separate set of experiments with varied ion energies \Heplus ions from 1 to 5 keV were used. Typical ion current during analysis was 1 - 4 nA. Whenever ion sputtering was performed, a separate ion gun with 0.5 keV Ar\supt{+} ions at angle of incidence of 59\degree and ion current of 100 nA was used.

To obtain the deposition depth profiles for Ru on B, \BfC and C a series of samples with a varied thickness of Ru layer was produced for each material combination. The general composition of samples was the following: the initial substrate was a superpolished Si wafer covered with its native oxide, then it was covered with 4 nm of amorphous Si, then 5 nm of the chosen substrate layer (B, C or \BfC), then 0 -- 35 nm of Ru. Immediately after deposition each sample was transferred to the LEIS chamber and analyzed, the delay between deposition and analysis was within 10 -- 15 minutes. The results of these measurements are presented below.

It is worth noting that we distinguish individual samples of each series by the as-deposited thickness of Ru film, which is obtained from calibrated signal of a quartz crystal microbalance. Due to interaction of Ru with its substrate the density of the grown film is different from a bulk Ru film used for calibration. Therefore, the as-deposited Ru thickness is not equal to the actual Ru film thickness, but instead serves as a measure of the deposited amount of Ru.

\section{LEIS measurements with 3 keV He+} \label{RLsec:LEIS_3keV}

\subsection{LEIS spectra} \label{RLsec:LEIS_spectra}

\begin{figure}[t]
    \centering
    \begin{subfigure}[b]{0.475\textwidth}
        \centering
        \includegraphics[width=\textwidth]{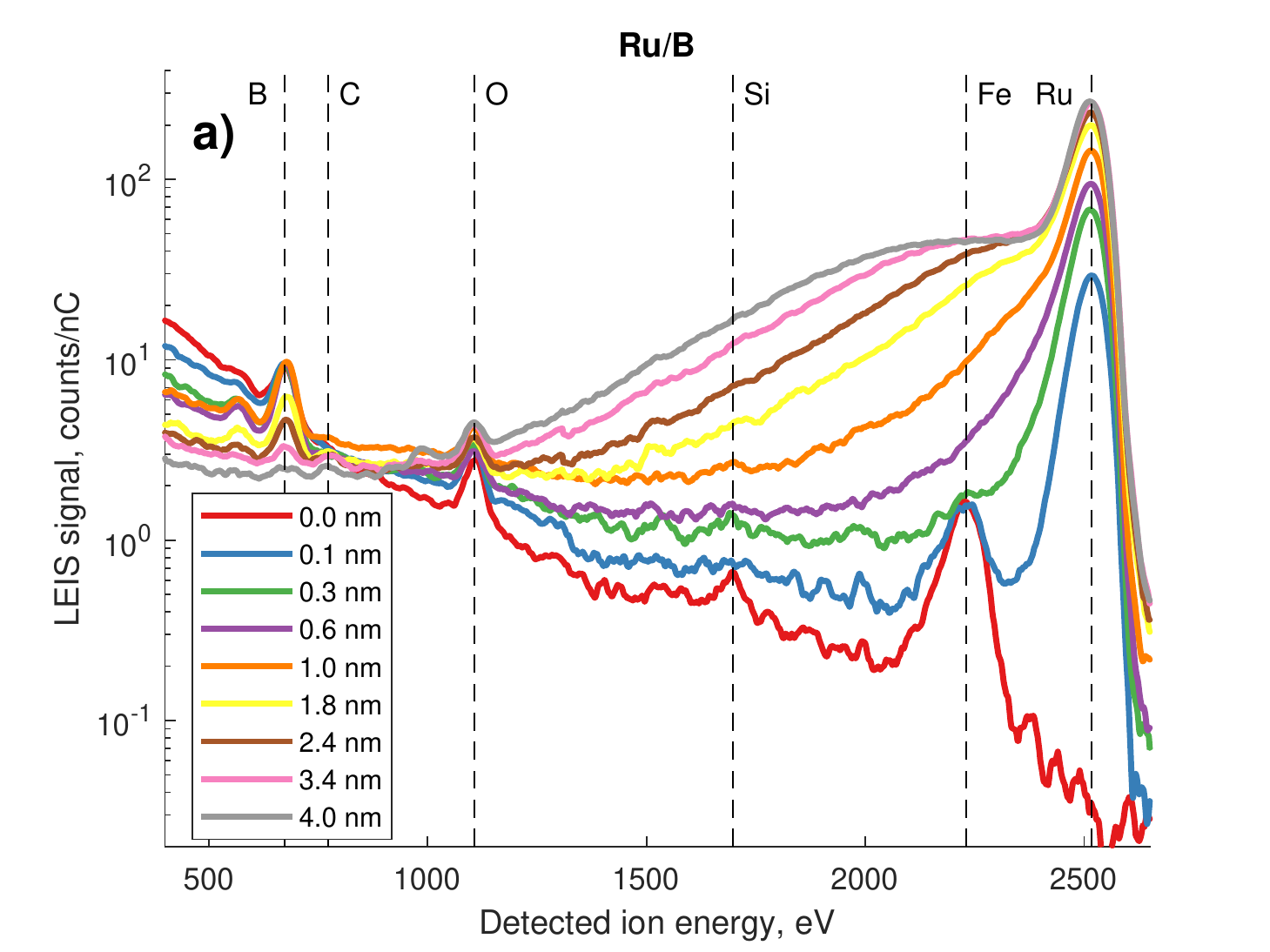}
        \phantomcaption    
        \label{RLfig:LEIS_spectra_Ru_B}
    \end{subfigure}
    \hfill
    \begin{subfigure}[b]{0.475\textwidth}  
        \centering 
        \includegraphics[width=\textwidth]{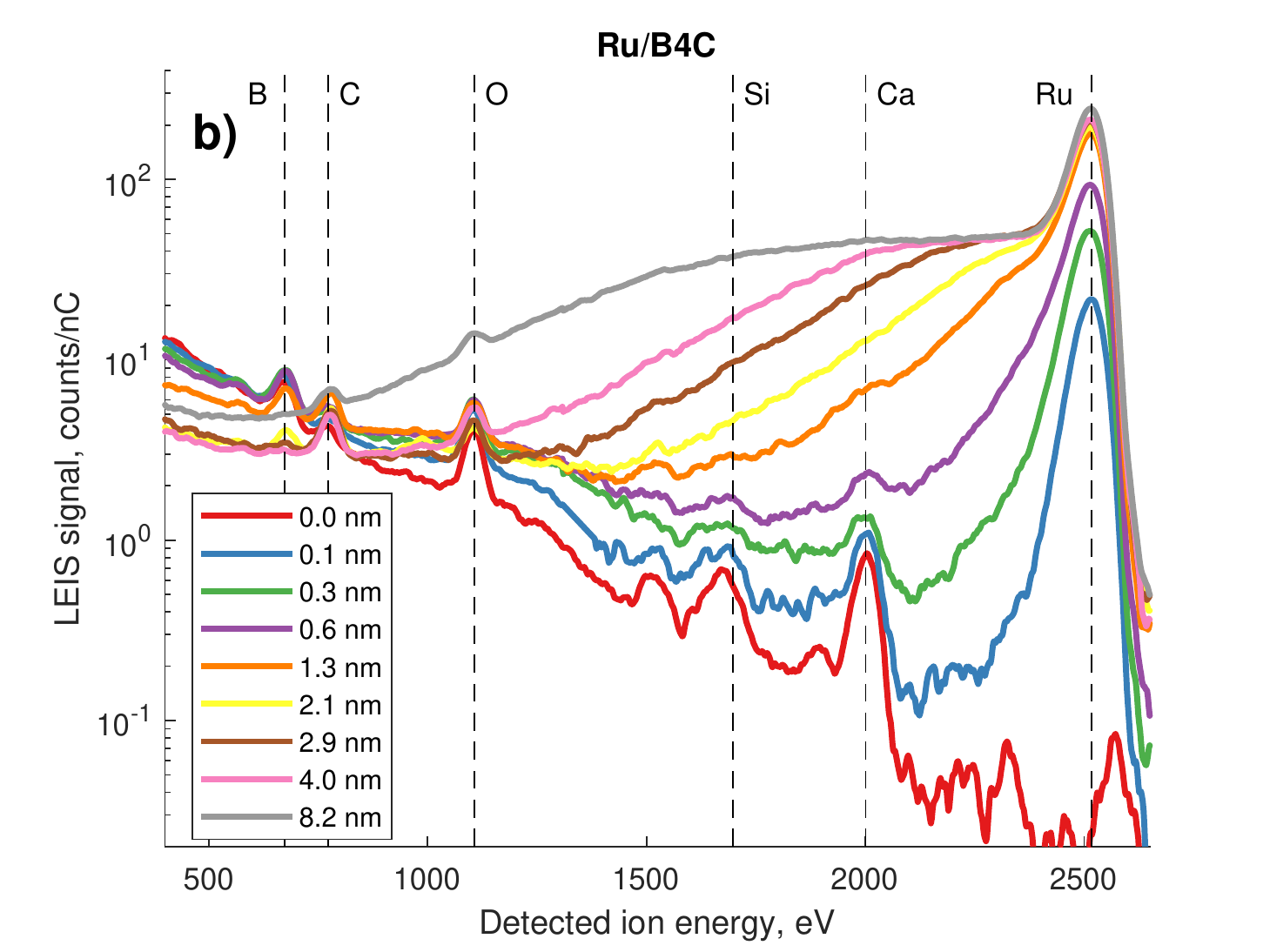}
        \phantomcaption    
        \label{RLfig:LEIS_spectra_Ru_B4C}
    \end{subfigure}
    \vskip\baselineskip
    \begin{subfigure}[b]{0.475\textwidth}   
        \centering 
        \includegraphics[width=\textwidth]{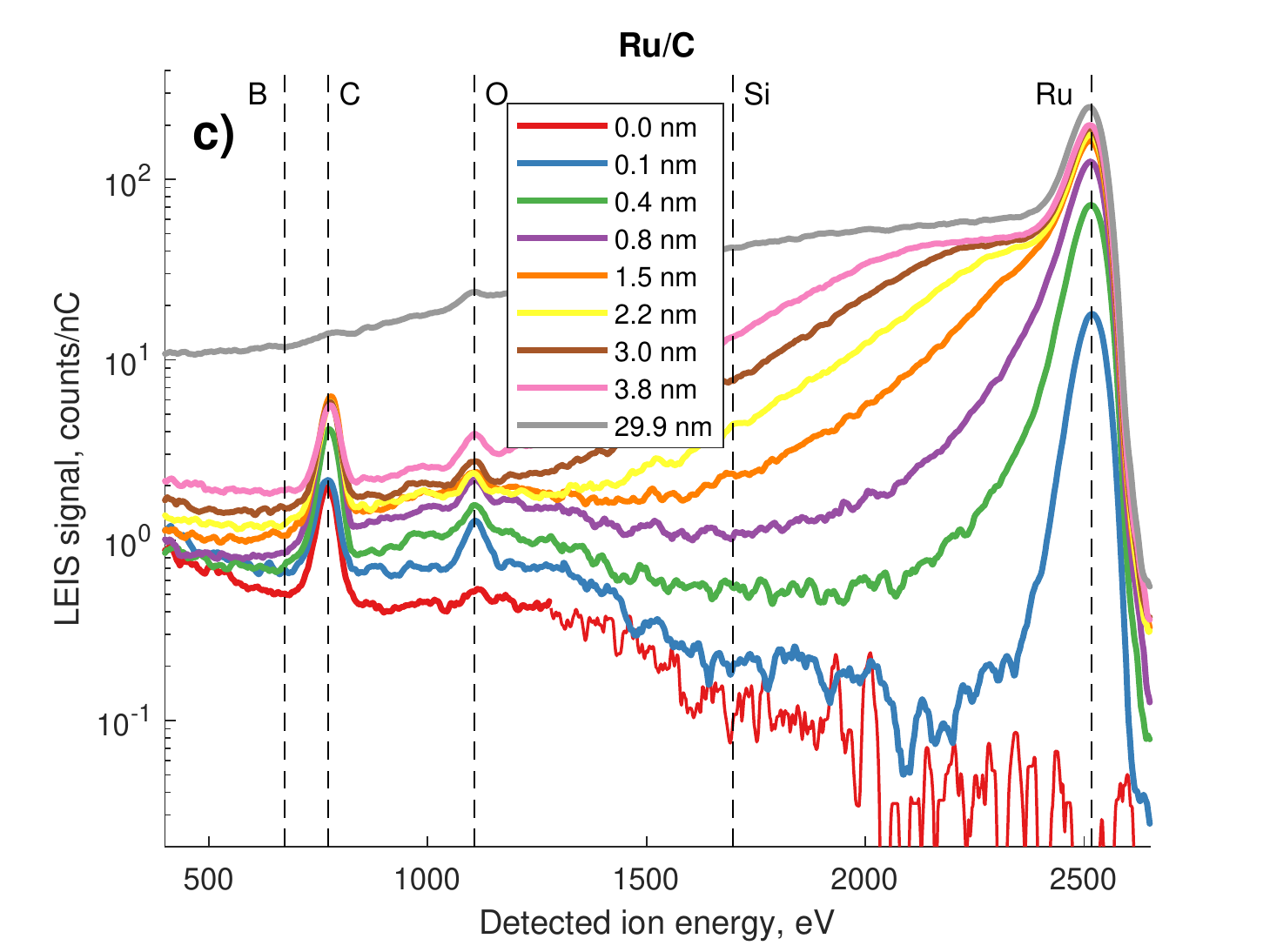}
        \phantomcaption    
        \label{RLfig:LEIS_spectra_Ru_C}
    \end{subfigure}
    \caption
    {3 keV \Heplus LEIS spectra of Ru thin films with different Ru thicknesses grown on B, \BfC and C substrates (subfigures A, B, C). A logarithmic scale is used for the signal axis. Surface backscattering peaks for each observed element (B, C, O, Si, Ca, Fe, Ru) are labelled according to experimentally observed peak energies. Subsurface scattering from Ru atoms -- the Ru tail -- is present at the low energy side of the Ru surface peak in each set of spectra. The thicknesses of the top Ru films are shown to label the spectra.} 
    \label{RLfig:LEIS_spectra}
\end{figure}

Raw LEIS spectra of each material combination are shown in Fig.~\ref{RLfig:LEIS_spectra}. Each spectrum has elemental surface peaks corresponding to 3 keV \Heplus backscattering from atoms of each element in the first atomic layer. Samples with Ru have a Ru ``tail'' at energies below the Ru peak, which represents scattering from Ru atoms located in deeper layers, followed by reionisation when the scattered particle leaves the surface.

Except for three elements deposited (Ru, B, C), there are other elements observed on the surface. A small oxygen peak is always present in the spectra in roughly the same amount, suggesting a slight surface contamination during sample transfer time. \BfC films have a negligible amount of Ca contamination, which may be present due to machining of the \BfC target and is only visible on the surface of \BfC due to the low surface energy of Ca~\cite{Mezey1982}. Boron substrate films have some Fe contamination, which can be expected to happen in a stainless steel chamber with a stainless steel magnetron chimney due to the non-localized nature of the plasma during RF magnetron sputtering. Finally, most spectra feature a background exponentially rising at lower energies, which consists of ions that were sputtered from the sample surface by \Heplus ions. 

\subsection{Surface peaks in LEIS spectra} \label{RLsec:Surface_peaks}

The integral area of a surface peak of an element i, expressed in counts per primary ion dose (here counts/nC), is given by the following equation~\cite{Brongersma2007}:

\begin{equation}
    S_i = \frac{1}{e} \xi RP_i^+  \frac{d\sigma_i}{d\Omega} N_i,
    \label{RLeq:LEIS_signal_equation}  
\end{equation}

where $e$ is the electron charge, $\xi$ is the analyzer and detector instrumental factor, $R$ is a roughness factor, $P_i^+$ is the ion fraction, $\frac{d\sigma_i}{d\Omega}$ is the differential cross-section in area per solid angle and $N_i$ is the surface atomic density in (atoms/area). The quantification of surface composition is simple when most of these parameters are constant or known. In the ideal case, we expect $S_i \propto N_i$. Such proportionality is obtained under the following assumptions. First, $\xi = \mathrm{const}$, because the same setup is used for the measurements, and analyzer and detector configurations are fixed. Second, $\frac{d\sigma_i}{d\Omega} = \mathrm{const}$ and only depends on the ion-target atom combination, scattering geometry and ion energy. Third, as long as our thin films do not exhibit island growth, and we mostly deal with amorphous surfaces, we can assume $R = \mathrm{const}$. This factor is most probably (slightly) varying during growth, but without additional in-situ information of high frequency roughness development we have assumed a constant roughness. Finally, in the absence of matrix effects $P_i^+ = \mathrm{const}$ if measurements are done at the same energy. The involvement of matrix effects will be a subject of the Section~\ref{RLsec:varE_measurements}.

When matrix effects are absent, the incident ion energies are fixed and roughness is constant, Eq.~\ref{RLeq:LEIS_signal_equation} becomes a simple proportionality $S_i \propto N_i$. Following Brongersma et. al.~\cite{Brongersma2007}, we convert $N_i$ into surface coverage ${\vartheta}_i=\frac{N_i}{N_i^{ref}}=\frac{S_i}{S_i^{ref}}$. Another surface quantity is the surface atomic fraction $x_i = \frac{N_i}{\sum_j N_j}$. The surface atomic fraction $x_i$ is a more representative quantity than surface coverage ${\vartheta}_i$, because $\sum_i x_i =1$ by definition, while $\sum_i {\vartheta}_i =1$ only in absence of compaction/expansion during compound formation. 

To calculate $N_i$, $\vartheta_i$ and $x_i$ we need $S_i^{\mathrm{ref}}$ and $N_i^{\mathrm{ref}}$ of every element present in the spectra. Reference samples with sputter cleaned surfaces of Ru, B and C were prepared to obtain $S_{\mathrm{Ru}}^{\mathrm{ref}}$, $S_{\mathrm{B}}^{\mathrm{ref}}$ and $S_{\mathrm{C}}^{\mathrm{ref}}$. Using this method, we obtain $S_{\mathrm{Ru}}^{\mathrm{ref}}=23600$ counts/nC, $S_{\mathrm{B}}^{\mathrm{ref}}=356$ counts/nC and $S_{\mathrm{C}}^{\mathrm{ref}}=76$ counts/nC. However, we will see in the next paragraph that there are problems with the $S_i^{\mathrm{ref}}$ obtained this way. The values of $S_{\mathrm{O}}^{\mathrm{ref}}$ and $N_{\mathrm{O}}^{\mathrm{ref}}$ were taken from reference~\cite{ColomaRibera2016}. Surface atomic densities $N_i^{\mathrm{ref}}$ for other elements were taken as 95\% of their tabulated bulk values to account for the less dense films formed by magnetron sputtering.

\begin{figure}[t]
    \centering
    \includegraphics[width=0.7\linewidth]{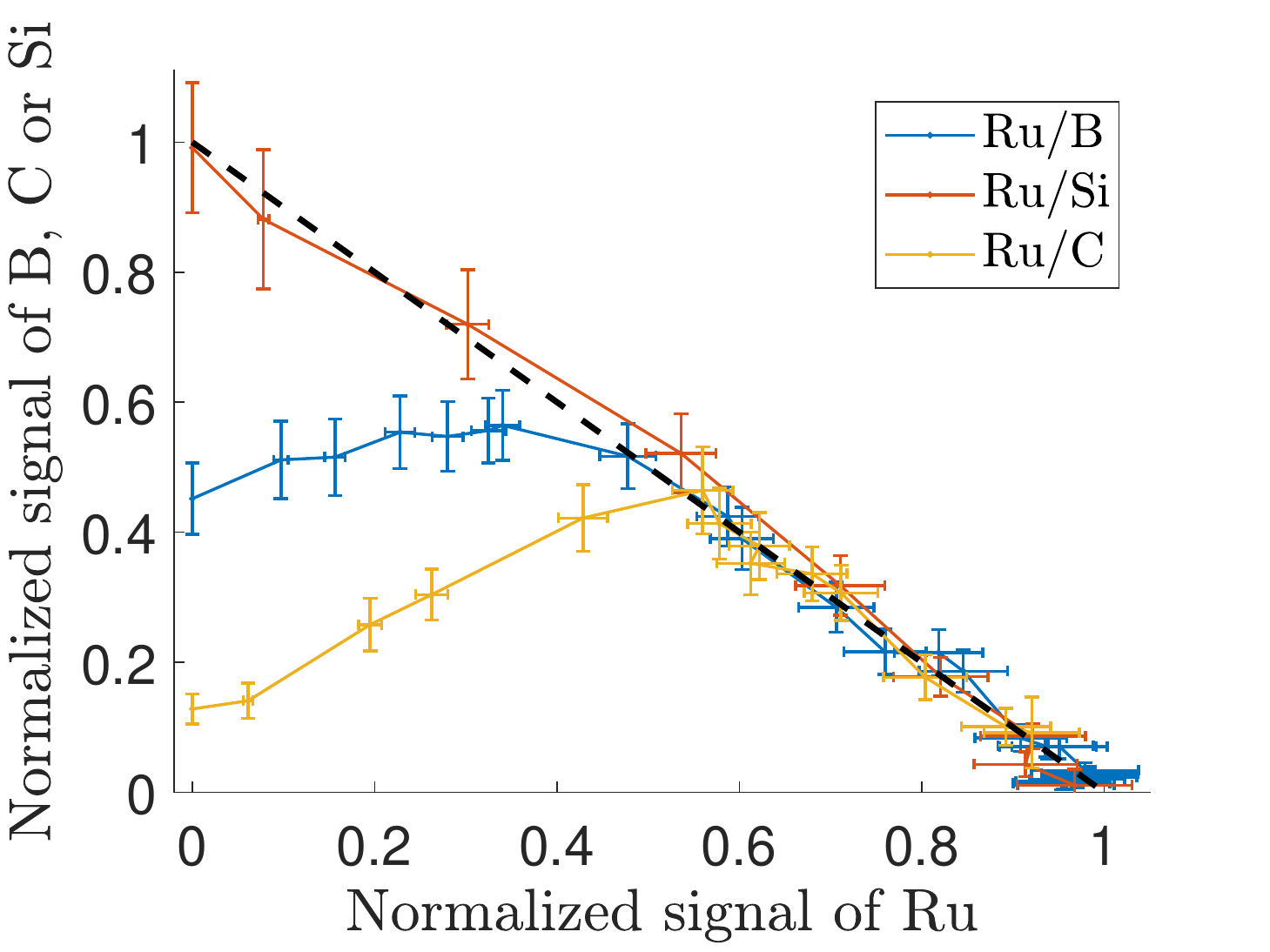}
    \caption{Pairwise comparison of normalized signals of elements in each binary system: Ru/B, Ru/C, Ru/Si. Such plot is also known as the ``matrix effect check''. For each point not subjected to a matrix effect the normalized signal is equal to the surface coverage of the respective element. Note that signal is normalized using the reference signals already obtained through this graph.}
    \label{RLfig:matrix_effect_check}
\end{figure}

Before calculation of $N_i$, $\vartheta_i$ and $x_i$ we can perform an internal calibration of $S_{\mathrm{Ru}}^{\mathrm{ref}}$, $S_{\mathrm{C}}^{\mathrm{ref}}$ and $S_{\mathrm{B}}^{\mathrm{ref}}$ with the help of a so-called ``matrix effect check'', in which two LEIS signals are plotted against each other. If we ignore a minor contribution of O contamination, Ru-C and Ru-B become binary systems, and in the absence of matrix effects and compaction/expansion we expect $\vartheta_{\mathrm{Ru}}+\vartheta_{\mathrm{C}}=1$ and $\vartheta_{\mathrm{Ru}}+\vartheta_{\mathrm{B}}=1$. ``Matrix effect checks'' for the Ru-C and Ru-B pairs are shown in Fig.~\ref{RLfig:matrix_effect_check}, where they are compared with the ``matrix effect check'' for the Ru-Si pair~\cite{ColomaRibera2016}. Ru-Si pair exhibits a behavior very close to linear: $\vartheta_{\mathrm{Ru}}+\vartheta_{\mathrm{Si}}=1$. The Ru-C pair exhibits more complicated behavior: at low Ru coverages both Ru and C signals increase together, but after approximately 50\% Ru coverage it switches to the same linear dependence as the Ru-Si pair. The behavior of the Ru-B pair has similarity with the Ru-C pair. 

A strong non-linear behavior suggests the presence of a matrix effect. The proper way to study matrix effects involves measuring LEIS signals with different ion energies, which will be done in Section~\ref{RLsec:varE_measurements}. Before the detailed study we can already assume that there are two different regimes of behavior of Ru-C and Ru-B curves in Fig.~\ref{RLfig:matrix_effect_check}, separated by certain critical amount of Ru $\vartheta_{\mathrm{Ru}} \approx 0.5$, achieved at $\thicksim 1$ nm of Ru. Any measurement that falls on the left side of the curve is suspected to have a matrix effect, and any measurement on the right side follows a straight line, which is expected in absence of matrix effects. If this assumption is correct, the right side of the curve can be used for internal calibration of $S_{\mathrm{Ru}}^{\mathrm{ref}}$ and $S_{\mathrm{C}}^{\mathrm{ref}}$ when not affected by the matrix effect. 

For the Ru-C pair, the extrapolation gives $S_{\mathrm{Ru}}^{\mathrm{ref}} = 21600$ counts/nC and $S_{\mathrm{C}}^{\mathrm{ref}} = 592$ counts/nC. The value for Ru is not very different from a clean sputtered reference, however for C the difference is almost 8 times. These references values can be used to quantify the linear part of the Ru-C binary system. For the Ru-B pair we extract $S_{\mathrm{Ru}}^{\mathrm{ref}} = 21970$ counts/nC and $S_{\mathrm{B}}^{\mathrm{ref}} = 738$ counts/nC. The new reference value for B is two times higher than the original. For Ru/\BfC samples we use the reference signals of Ru, B and C as for Ru/B and Ru/C separately.

\subsection{Deposition depth profiles} \label{RLsec:Depo_DPs}

\begin{figure}[!t]
    \centering
    \begin{subfigure}[b]{0.475\textwidth}
        \centering
        \includegraphics[width=\textwidth]{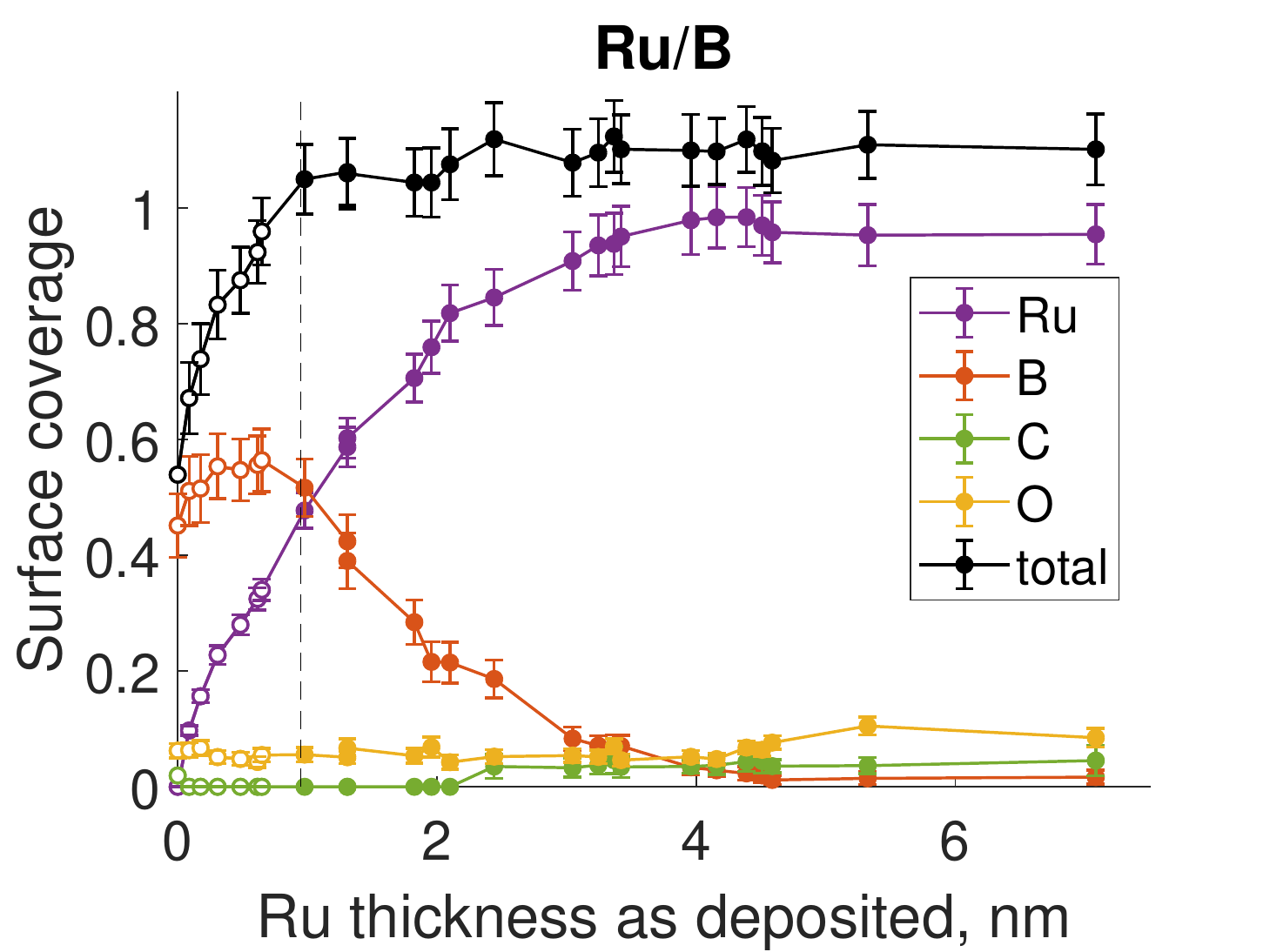}
        \phantomcaption    
        \label{RLfig:LEIS_depoDP_Ru_B_sc}
    \end{subfigure}
    \hfill
    \begin{subfigure}[b]{0.475\textwidth}  
        \centering 
        \includegraphics[width=\textwidth]{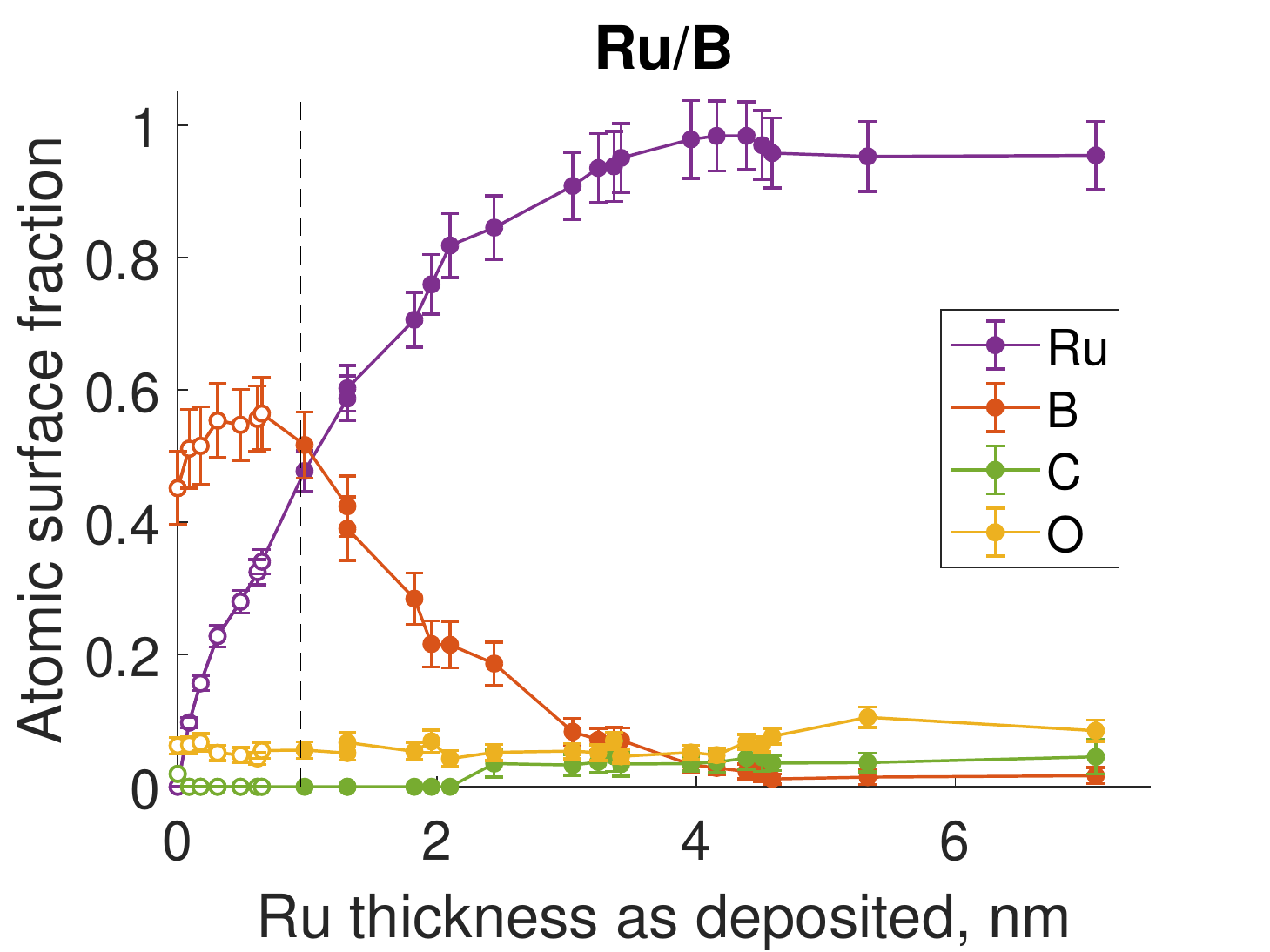}
        \phantomcaption    
        \label{RLfig:LEIS_depoDP_Ru_B_af}
    \end{subfigure}
    \vskip\baselineskip
    \begin{subfigure}[b]{0.475\textwidth}
        \centering
        \includegraphics[width=\textwidth]{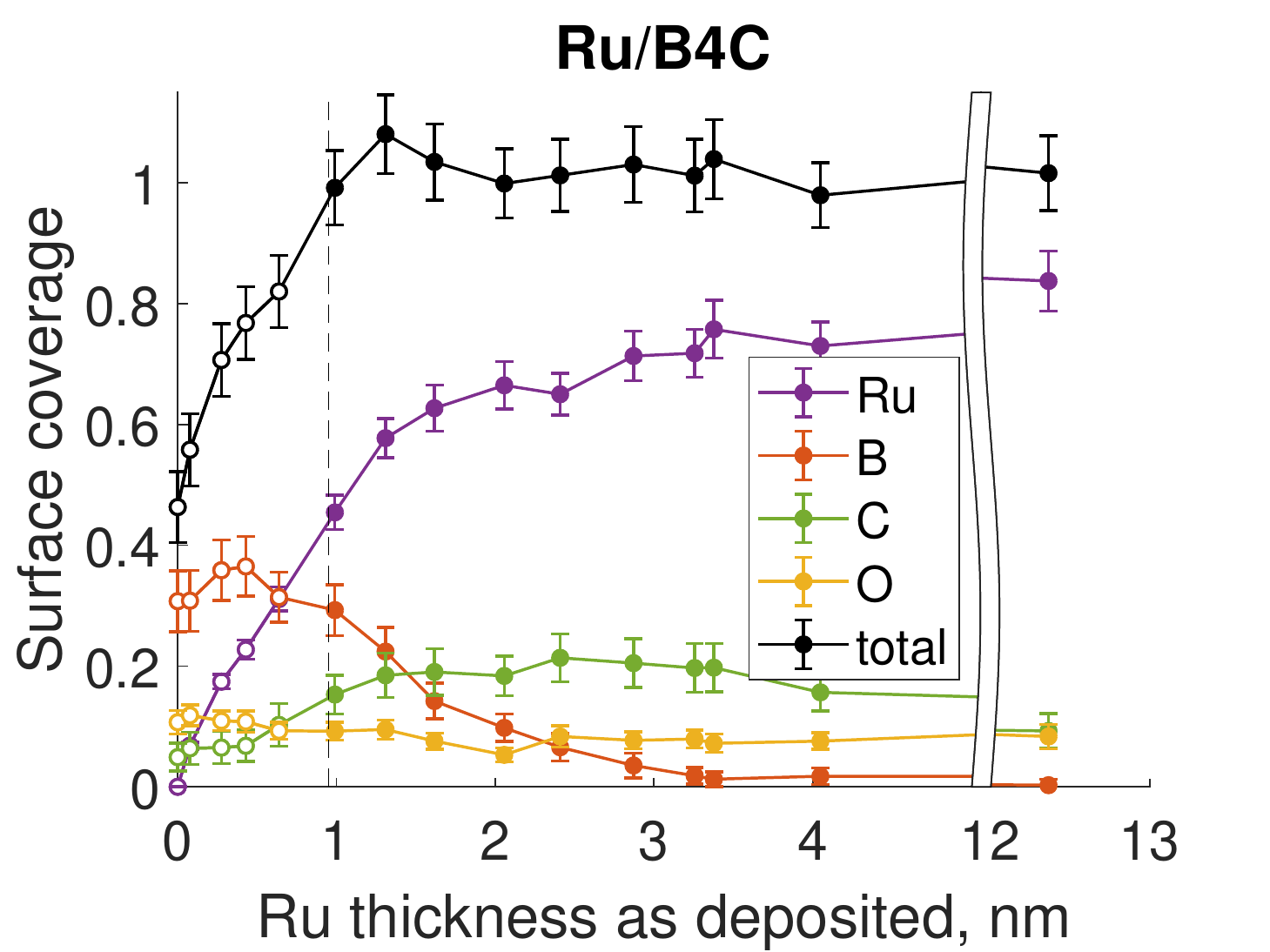}
        \phantomcaption    
        \label{RLfig:LEIS_depoDP_Ru_B4C_sc}
    \end{subfigure}
    \hfill
    \begin{subfigure}[b]{0.475\textwidth}  
        \centering 
        \includegraphics[width=\textwidth]{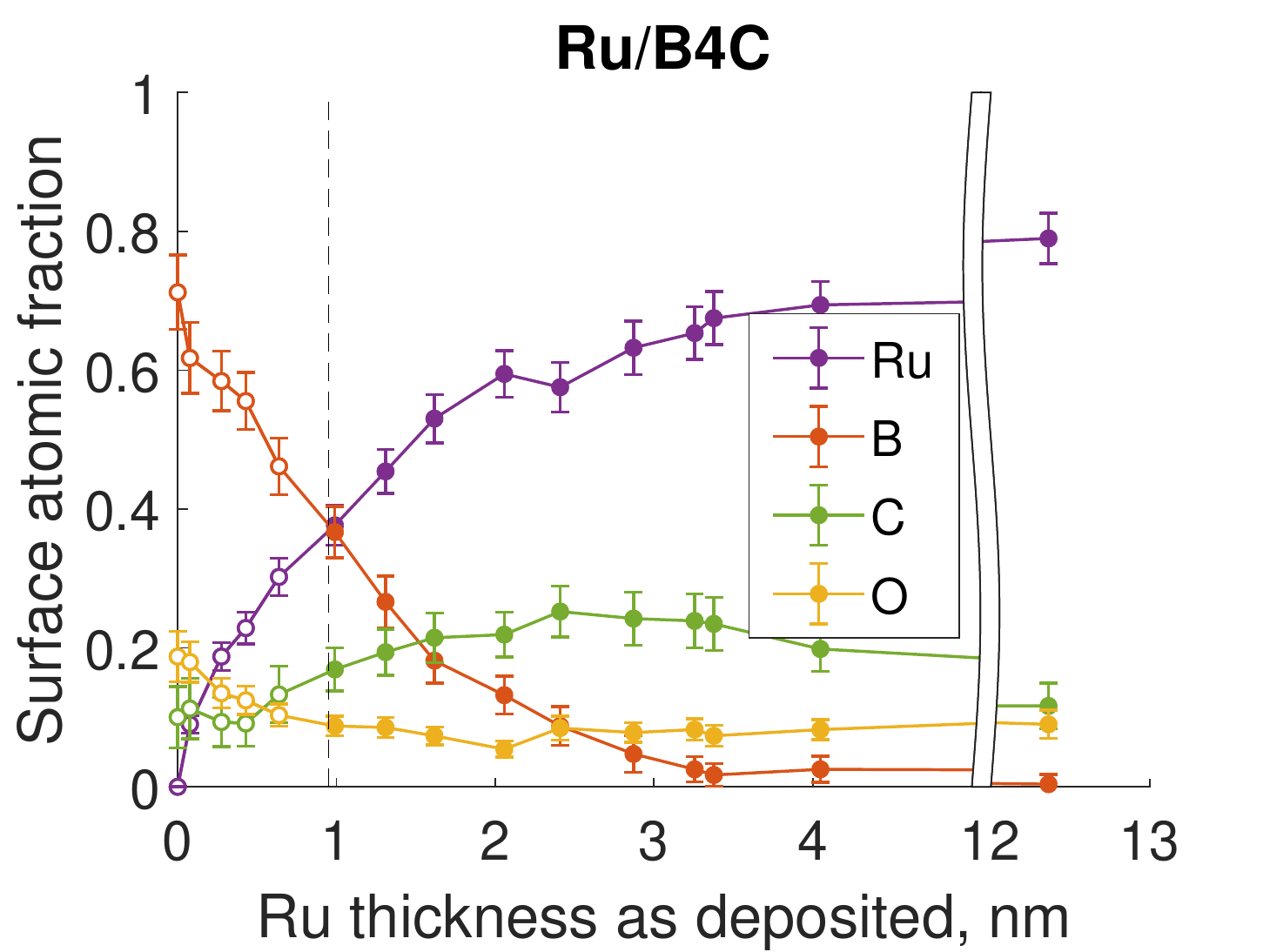}
        \phantomcaption    
        \label{RLfig:LEIS_depoDP_Ru_B4C_af}
    \end{subfigure}
    \vskip\baselineskip
    \begin{subfigure}[b]{0.475\textwidth}
        \centering
        \includegraphics[width=\textwidth]{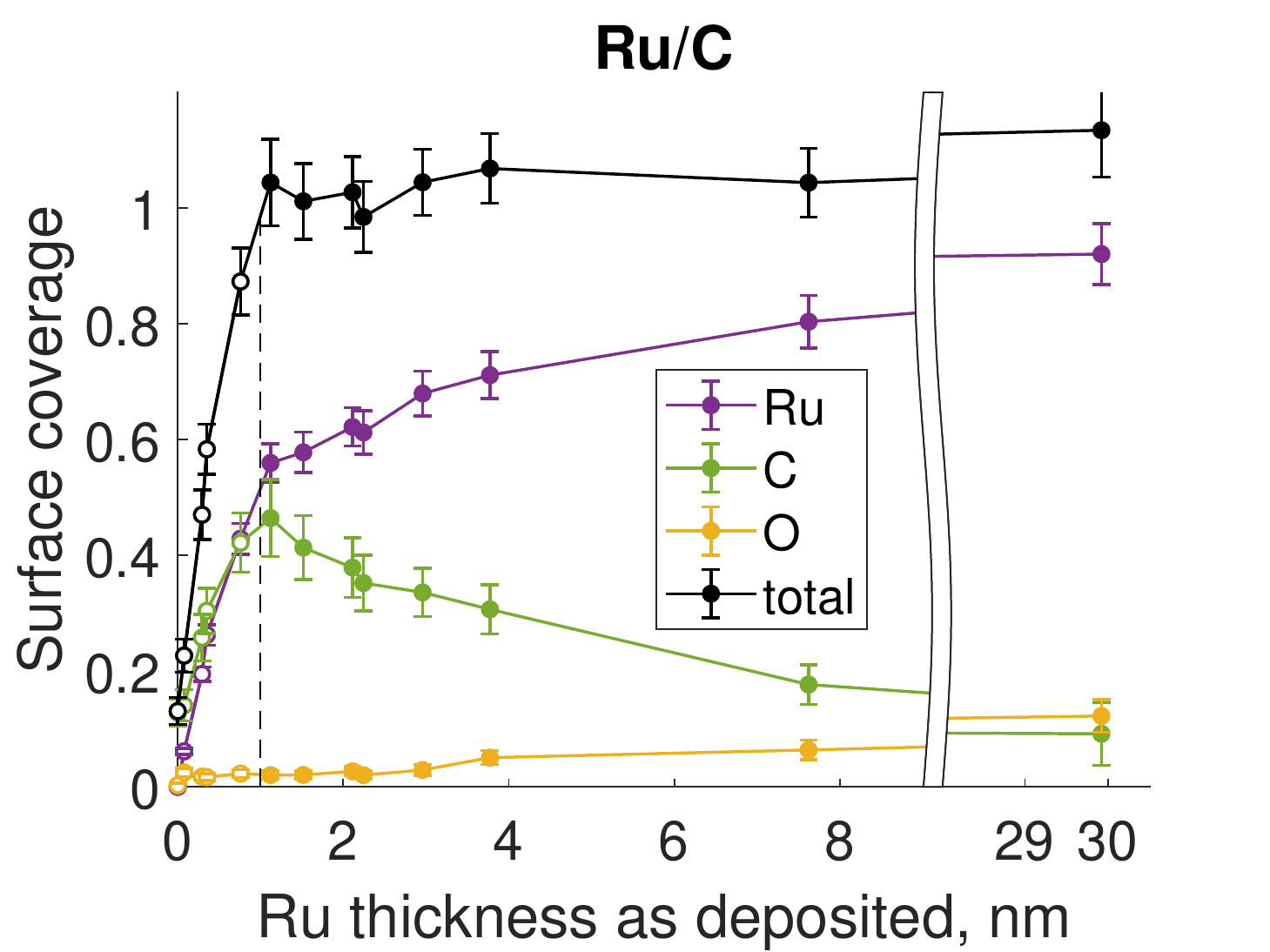}
        \phantomcaption    
        \label{RLfig:LEIS_depoDP_Ru_C_sc}
    \end{subfigure}
    \hfill
    \begin{subfigure}[b]{0.475\textwidth}  
        \centering 
        \includegraphics[width=\textwidth]{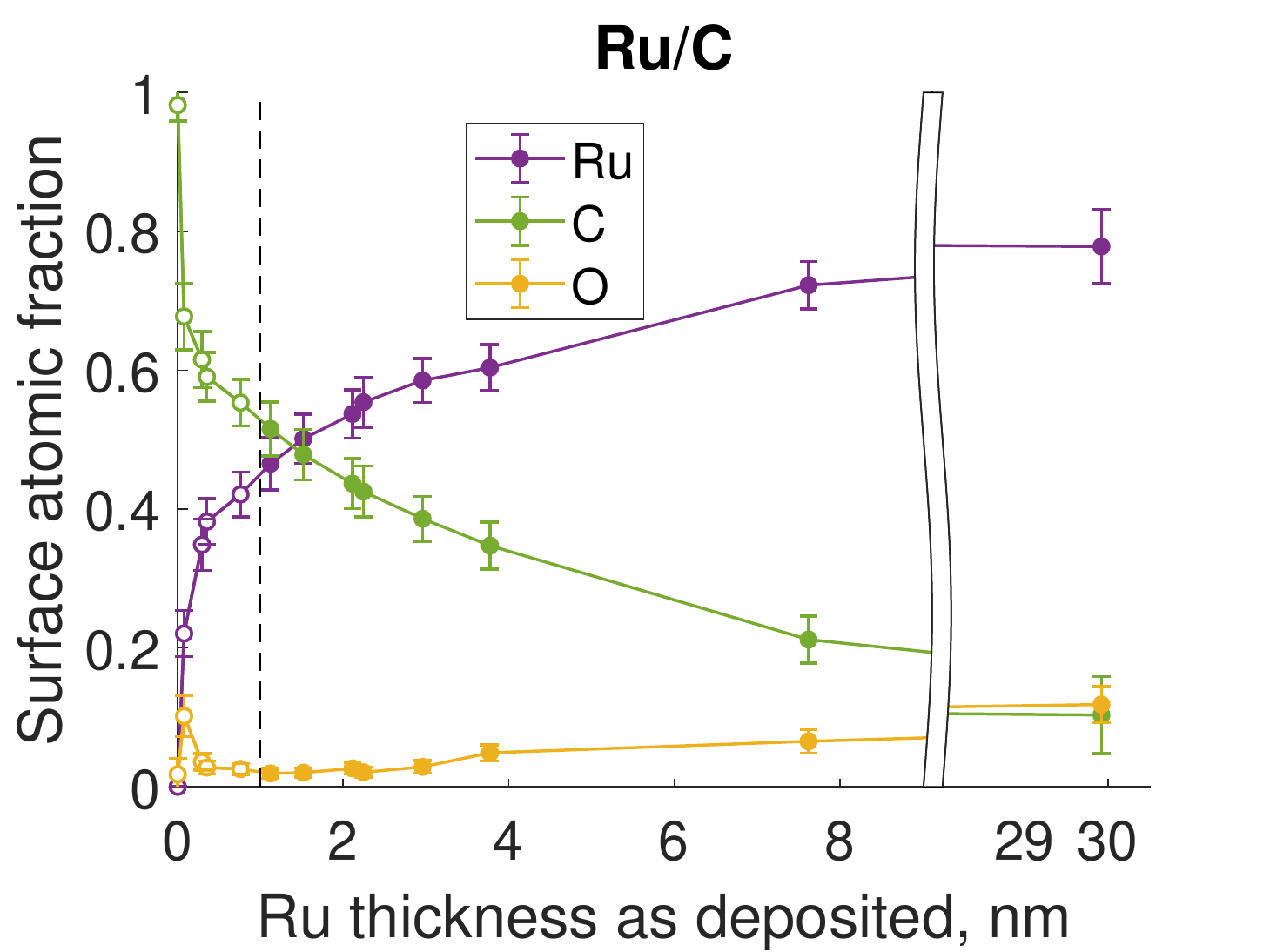}
        \phantomcaption    
        \label{RLfig:LEIS_depoDP_Ru_C_af}
    \end{subfigure}    
    \caption
    {Surface coverages (left side) and surface atomic fractions (right side) of Ru, B, C and O extracted from spectra in Fig.~\ref{RLfig:LEIS_spectra}. From top to bottom: Ru/ B, Ru/C, Ru/ \BfC. The data subjected to matrix effects are shown in open circles (to the left from the vertical dashed line) and are only presented to demonstrate the behavior of LEIS signals, but cannot be used for quantification.} 
    \label{RLfig:LEIS_depoDP}
\end{figure}

After the discussion of the aspects of quantification of the surface composition, we demonstrate the deposition depth profiles of Ru on different substrates (Fig.~\ref{RLfig:LEIS_depoDP}). Both surface coverages $\vartheta_i$ and surface atomic fractions $x_i$ for each element are presented. All data points that are subject to matrix effect are marked differently, and not used in the discussion. There are several observations to be made from these depth profiles. The fraction of Ru on the surface is monotonously increasing with the amount of Ru, the fraction of the substrate atoms is monotonously dropping -- B in Ru/B, C in Ru/C, fractions of B and C added together in Ru/\BfC. 

It is unexpected that these changes are occurring relatively slow -- in the worst case, i.e. in Ru/C, some carbon atoms are still visible after 30 nm of Ru is deposited on top of the C film. There are three possible explanations: it can be caused an extremely deep intermixing, growth of Ru in large islands, or surface segregation of C. Island growth is disproved by AFM measurements of the samples at different stages of growth. The root mean squared roughness was measured from 1.5 and 15 nm Ru films on C, B and \BfC. All of these samples exhibited stochastic roughness with an RMS value from 0.14 to 0.21 nm, similar to the initial substrate roughness, or to Ru grown on Si~\cite{ColomaRibera2016}. Therefore, island growth modes can be excluded. 

\begin{figure}[t]
    \centering
    \includegraphics[width=0.7\linewidth]{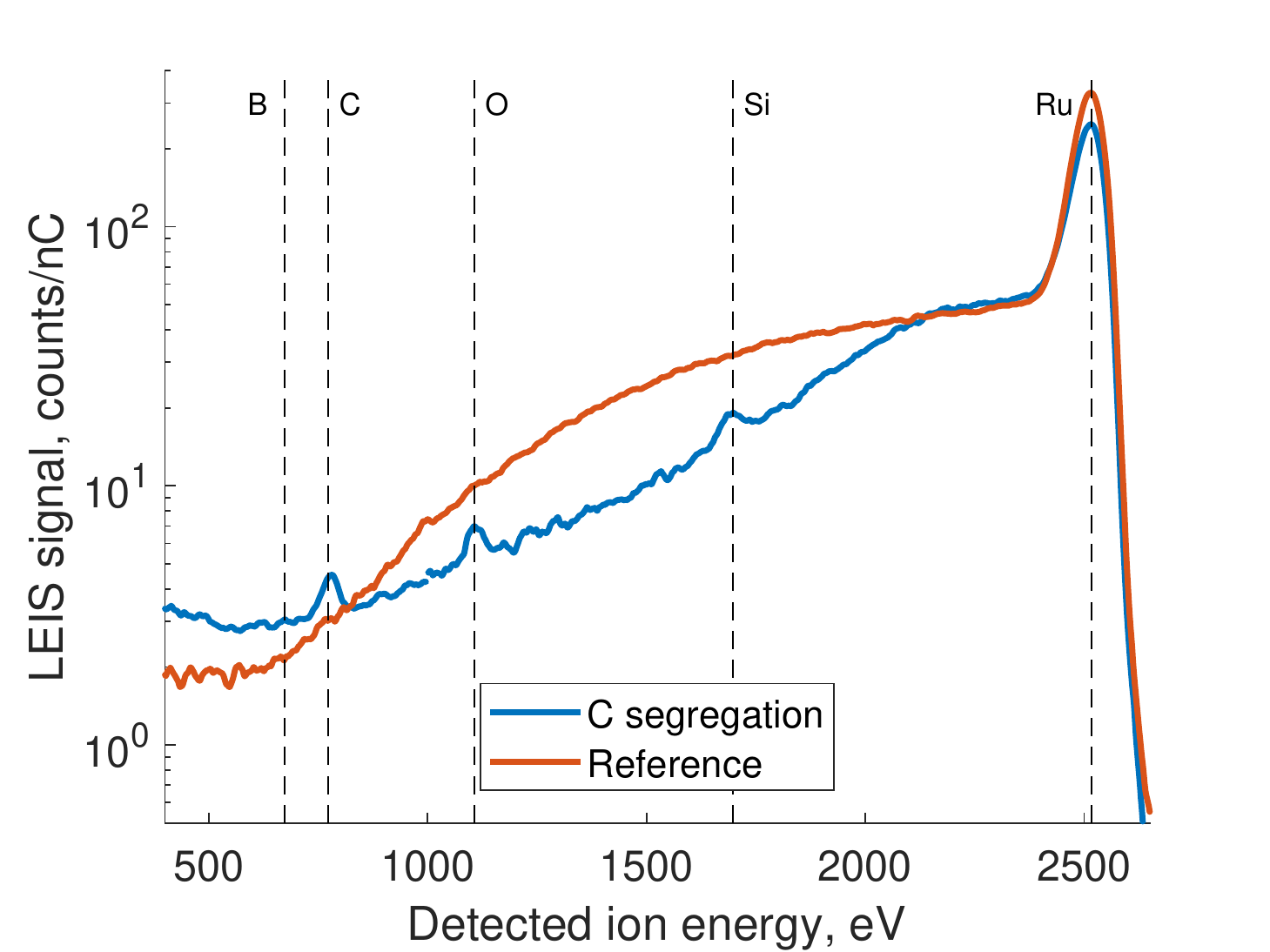}
    \caption{Comparison of a sample with pronounced surface segregation (blue curve) with a reference sample (red curve). The reference spectrum is taken from a sputter cleaned surface of a 14 nm Ru thin film grown on a 5 nm \BfC substrate on a Si substrate. C segregation spectrum is from an as-deposited surface of 3.2 nm Ru film on 0.4 nm \BfC layer on a Si substrate. This spectrum shows about 10\% C surface coverage, and therefore disproves the possibility of having a decaying or constant C content within the film.}
    \label{RLfig:LEIS_spectra_Ru_thin_B4C}
\end{figure}

To prove that a very slow disappearance of C from the surface cannot be explained by simple intermixing of C with Ru, we created a sample with a drastically reduced amount of C. A thickness of 3.2 nm of Ru was deposited on top of 0.4 nm of \BfC on a Si substrate. The LEIS spectrum of the surface of this sample is shown in Fig.~\ref{RLfig:LEIS_spectra_Ru_thin_B4C}. The surface coverage of C in this spectrum is about 10\%. A simple estimation shows that if all of the C atoms from the \BfC layer were spread homogenously in the Ru layer, it would result in 3\% volume concentration of C. This means that the C content within the Ru film cannot be constant or decaying toward the surface, but must increase at some point. Therefore, a surface segregation phenomenon is required to explain the observed effect.

The surface segregation of C on Ru is a well-known phenomenon in graphene growth~\cite{Wintterlin2009}, and we can apply it to Ru on C growth as well. Even after the interface transition from C to Ru is finished and the Ru layer becomes bulk, some C still continues to stay on the surface, only very slowly dissolving in the Ru matrix under the incoming flux of Ru atoms. Previous experiments~\cite{ColomaRibera2016} showed that Si is also segregating on top of Ru. We cannot exclude the possibility of surface segregation of B as well. 

Another notable feature, which we will call ``uplifting'', is related to the behavior of C is found in Ru/\BfC deposition depth profiles (Fig.~\ref{RLfig:LEIS_depoDP_Ru_B4C_af}). While the fraction of Ru increases and the fraction of B decreases, the fraction of C has a more complex behavior -- initially it increases, and only starts to drop when B disappears. This effect goes well beyond the thickness range involved in the matrix effect. We can assume that this feature is associated to surface segregation of C as well. If we assume that surface segregation only occurs on certain sites on a Ru surface, then the initial increase of the C signal is associated with the increase of the amount of the available sites. Upward diffusion of C atoms slows down with increasing thickness of the Ru layer, and after $\thicksim 2.5$ nm of Ru it is overcome by the process of removal of already segregated C atoms by arriving flux of fresh Ru atoms. 

Due to the fact that B and C fractions behave independently in Fig.~\ref{RLfig:LEIS_depoDP_Ru_B4C_af}, we conclude that \BfC is decomposing during the deposition of Ru. Decomposition of \BfC was already shown by XPS studies of this layer system~\cite{Tsarfati2009}. 

The last observation from Fig.~\ref{RLfig:LEIS_depoDP_Ru_B_af} is about the presence of C in the Ru/B deposition depth profile. In a pure B sample there is no detectable amount of C, but a small and nevertheless certain amount of C appears after 2 nm of Ru are deposited (also beyond the thickness range involved in the matrix effect). For comparison, no C signal was detected in Ru/Si growth~\cite{ColomaRibera2016}. The B target that was used for deposition always has 1-3\% C doping. This amount of C is undetectable in a pure B sample. Its appearance in later samples is a sign of surface segregation of C on Ru, which brings C atoms to the surface.

\section{LEIS measurements with different He+ energies} \label{RLsec:varE_measurements}

The matrix effects in Ru-B and Ru-C in Fig.~\ref{RLfig:matrix_effect_check} require additional attention. In this section we study the origin of matrix effects with focus on Ru-B material combination. We follow the established technique to study matrix effects in LEIS~\cite{Cortenraad2002,Zameshin2018a}, which we earlier used to study matrix effects in La-containing surfaces. The measure of neutralization efficiency is a so-called characteristic velocity $v_c$, which is in an integrated electron transfer rate from the surface to an ion. To obtain characteristic velocity for a given ion-target combination, LEIS signal needs to be measured at different incident ion energies and then plotted as a function of ion velocity in the special coordinate set given by Eq. (5) in~\cite{Zameshin2018a}.

\begin{figure}
    \centering
    \begin{subfigure}[b]{0.475\textwidth}
        \centering
        \includegraphics[width=\textwidth]{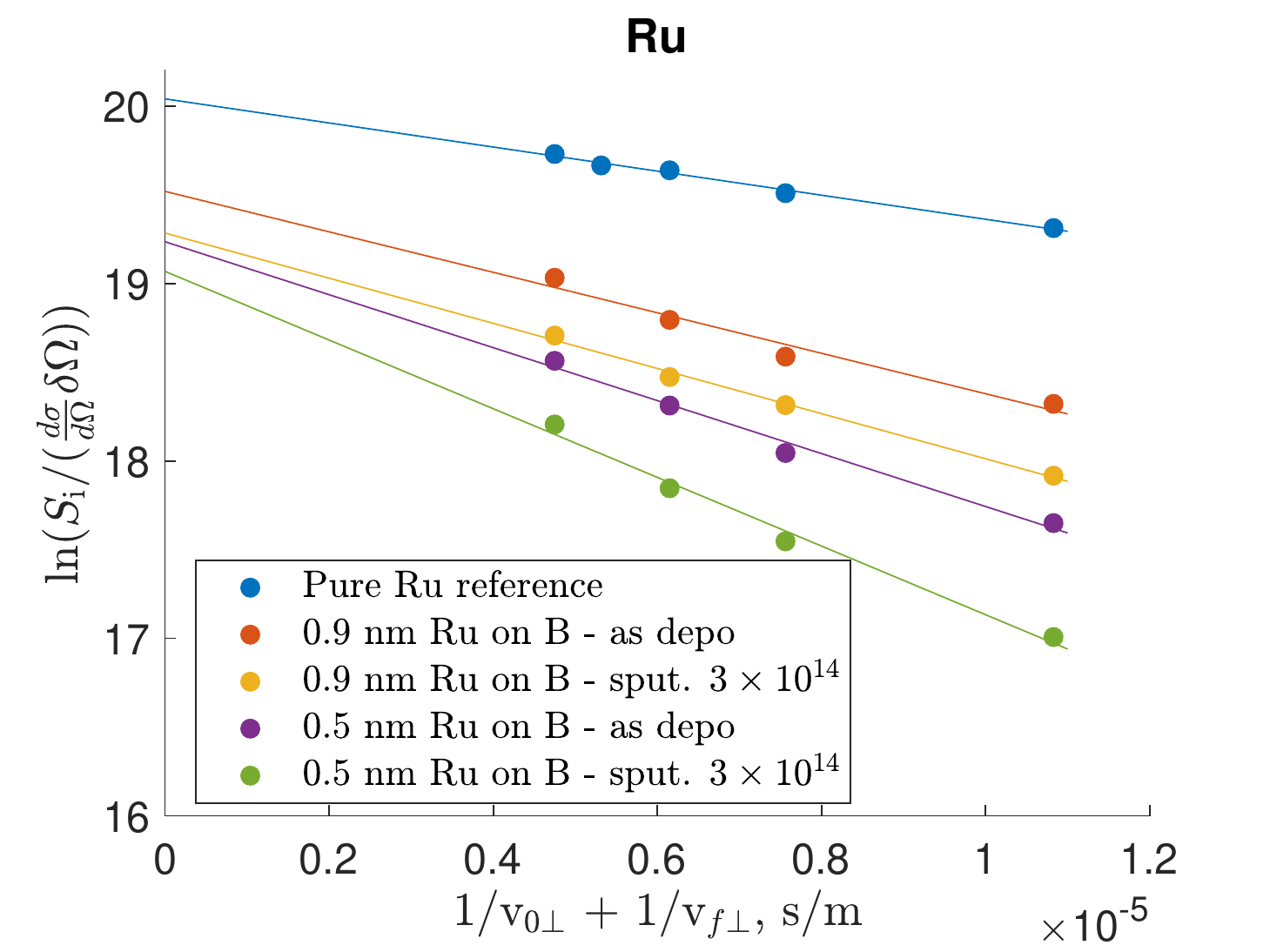}
        \phantomcaption    
        \label{RLfig:inv_velocity_plots_Ru}
    \end{subfigure}
    \hfill
    \begin{subfigure}[b]{0.475\textwidth}  
        \centering 
        \includegraphics[width=\textwidth]{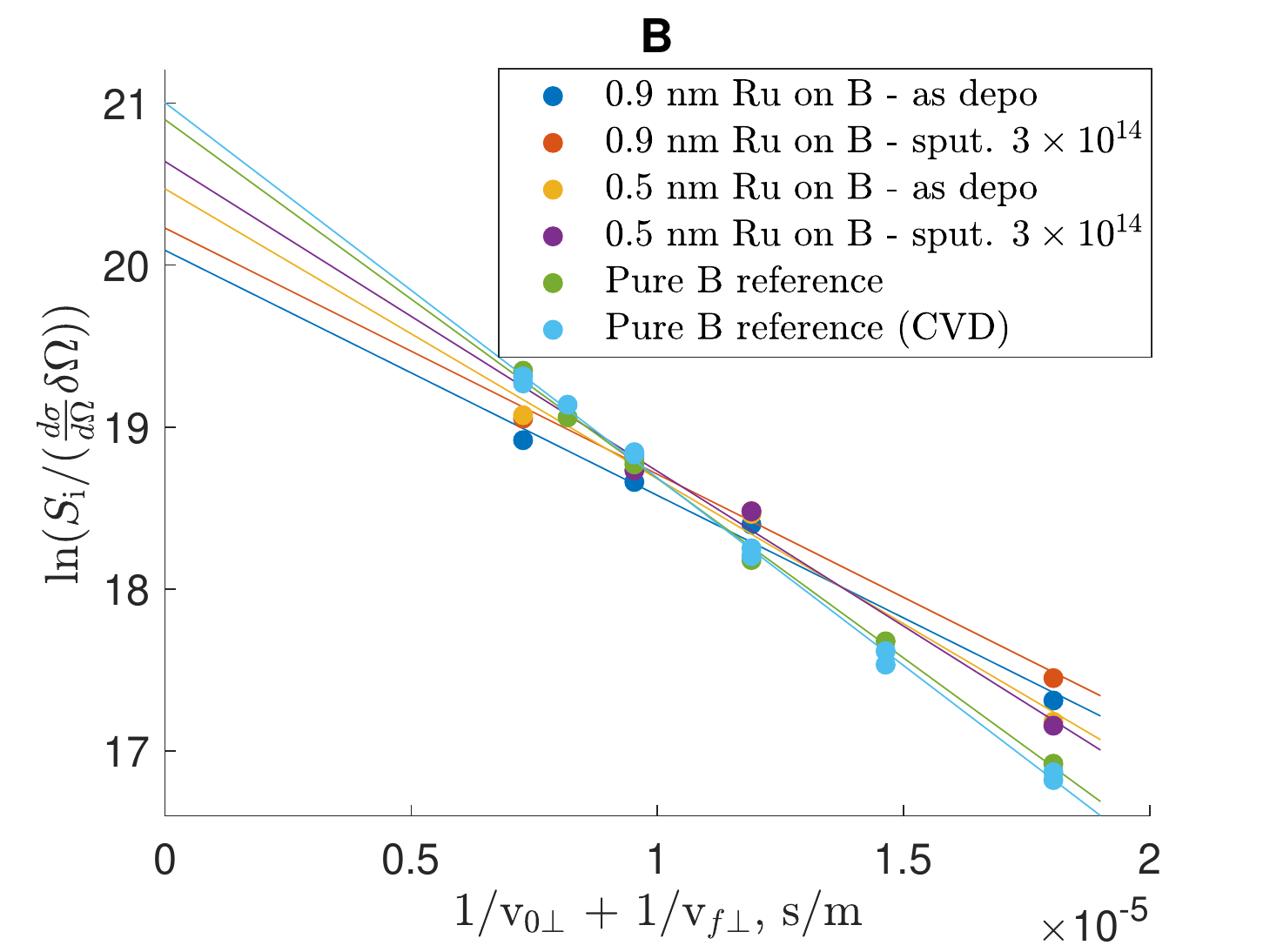}
        \phantomcaption   
        \label{RLfig:inv_velocity_plots_B}
    \end{subfigure}
    \caption
    {a) Dependence of logarithmic normalized LEIS signal of Ru in different samples on the inversed velocity of \Heplus ions. The least square fitted lines are shown for each set. The linear slope is $v_c$, the vertical offset is $\ln(C \times N_{\mathrm{Ru}} )$. b) The same for B.} 
    \label{RLfig:inv_velocity_plots}
\end{figure}

To measure characteristic velocity of pure Ru and pure B as reference samples, 20 nm Ru and 15 nm B thin films were deposited on Si wafers. LEIS signals of Ru or B were measured with several incident ion energies ranging from 1 to 5 keV. Before each measurement the surface was sputtered by a 0.5 keV Ar$^+$ beam until signal saturation. The data for Ru is shown in Fig.~\ref{RLfig:inv_velocity_plots_Ru} and the data for B is shown in Fig.~\ref{RLfig:inv_velocity_plots_B}. The next step was to perform similar measurements for the mixtures of Ru and B of different compositions, for which samples of 0.9 and 0.5 nm Ru on 5 nm B on Si wafer were chosen. To extend the range of measured surface compositions without depositing more samples, after the initial LEIS measurement each of these samples was sputtered with 0.5 keV Ar$^+$ ion beam of $3 \times 10^{14}$ atoms/cm\supt{2} fluence and measured again.

\begin{figure}
    \centering
    \begin{subfigure}[b]{0.475\textwidth}
        \centering
        \includegraphics[width=\textwidth]{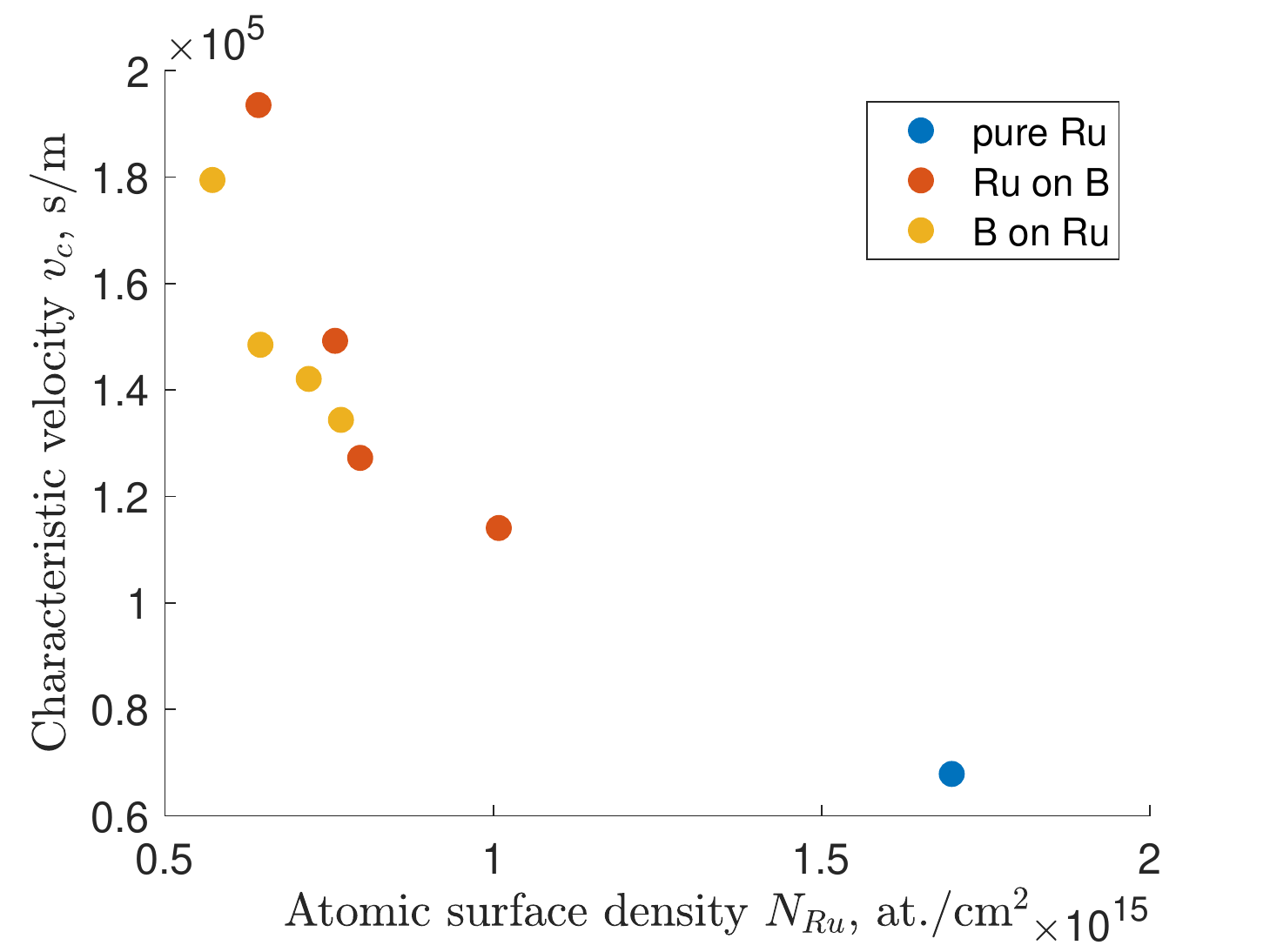}
        \phantomcaption    
        \label{RLfig:vc_vs_comp_Ru}
    \end{subfigure}
    \hfill
    \begin{subfigure}[b]{0.475\textwidth}  
        \centering 
        \includegraphics[width=\textwidth]{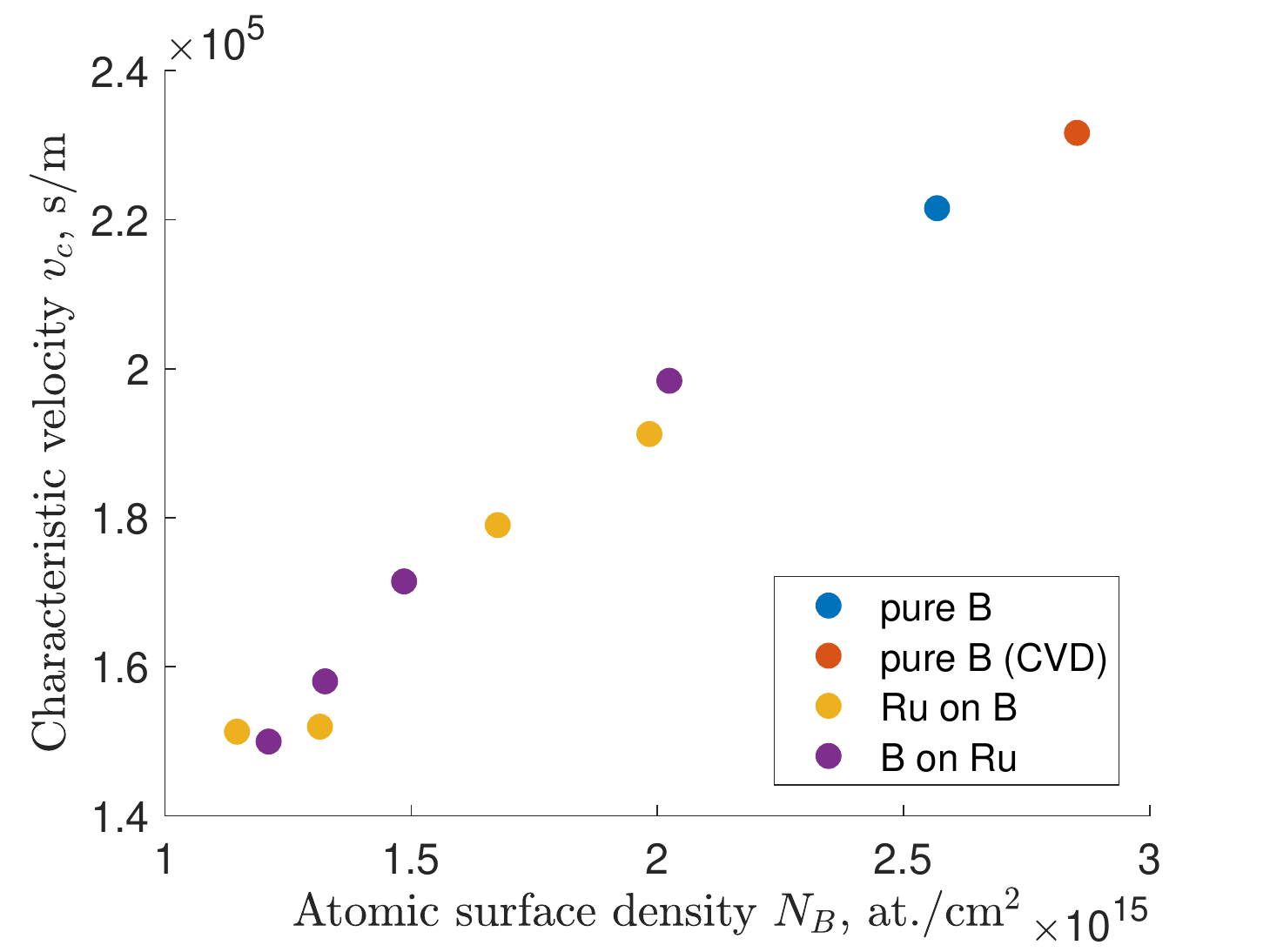}
        \phantomcaption   
        \label{RLfig:vc_vs_comp_B}
    \end{subfigure}
    \caption
    {a) Dependence of characteristic velocity of Ru $v_{c, \mathrm{Ru}}$ in different samples on the atomic surface density of Ru $N_{\mathrm{Ru}}$ in these samples. b) The same for B. }
    \label{RLfig:vc_vs_comp}
\end{figure}

The next step of data analysis is to extract characteristic velocities $v_{c, \mathrm{Ru}}$ and $v_{c, \mathrm{B}}$ from the slopes of linear fits of the inversed velocity plots and the atomic surface densities $N_{\mathrm{Ru}}$ and $N_B$ from the intersection points of the linear fits, following the same procedure as in~\cite{Zameshin2018a}. The results are shown in Fig.~\ref{RLfig:vc_vs_comp} separately for Ru and for B. Additional measurements were also performed on a reverse layer system as well: a sample of 0.5 nm B on 10 nm Ru was deposited and measured in the same fashion. To achieve different surface compositions for the reverse system, instead of a single sputter step, 3 sputter steps were done: with 1.5, 3 and 4.5 $\times 10^{14}$ atoms/cm\supt{2} fluences. The data for B/Ru films is shown in Fig.~\ref{RLfig:vc_vs_comp} as well. A simple conclusion from  Fig.~\ref{RLfig:vc_vs_comp} is that the characteristic velocities of both Ru and B strongly change with the surface composition regardless of the layer ordering. 
This proves the existence of matrix effect in the Ru-B material combination and associates it with the composition of the surface and not a specific atomic arrangement.

\begin{figure}[t]
    \centering
    \includegraphics[width=0.7\linewidth]{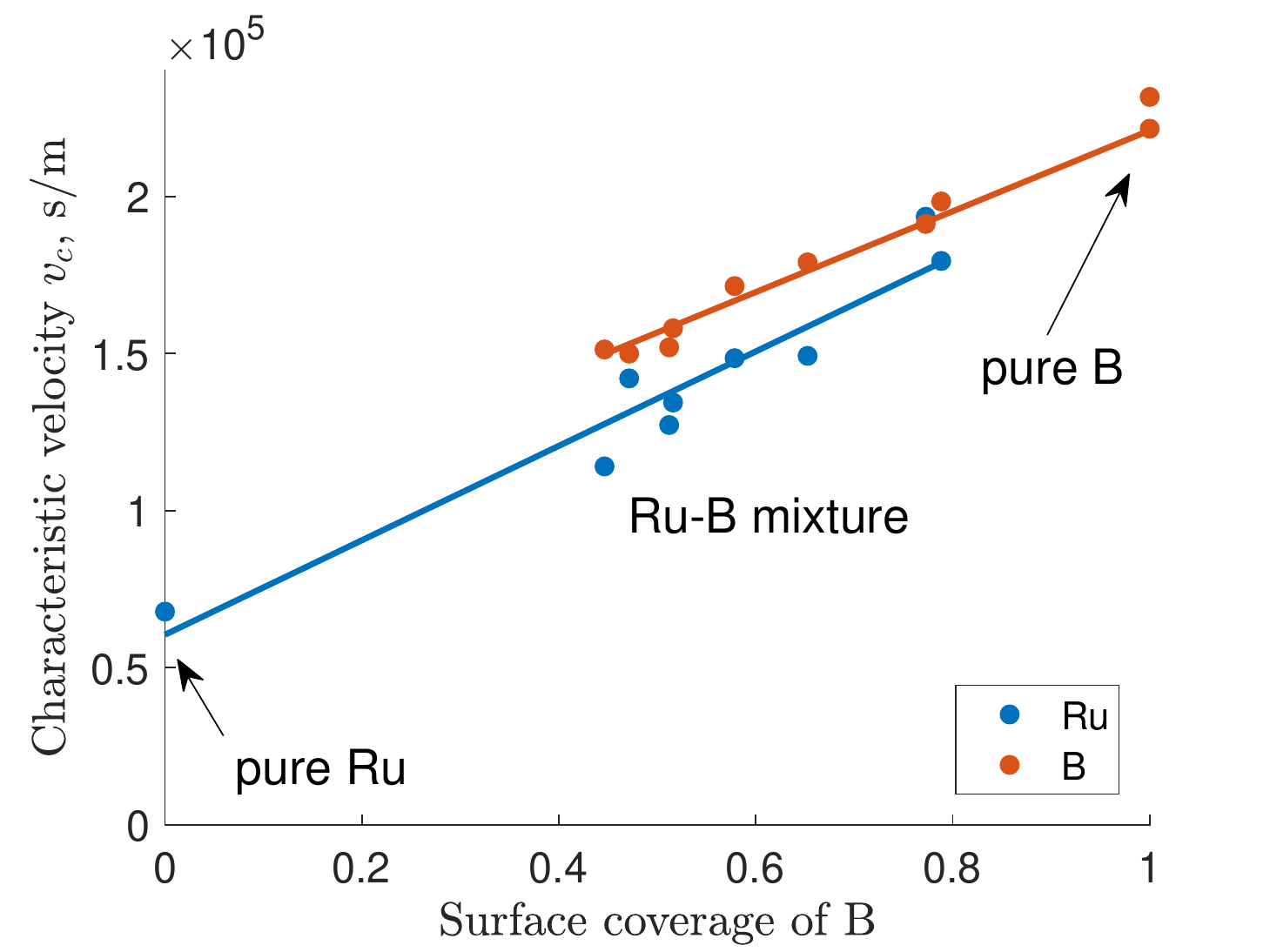}
    \caption{Linear behavior of characteristic velocities of Ru and B as a function of amount of B on the surface.}
    \label{RLfig:combined_vc_Ru_B_as_sc_B}
\end{figure}

To understand the mechanism of this matrix effect we need to study its behavior. $v_{c, \mathrm{B}}$ changes almost linearly with composition, while $v_{c, \mathrm{Ru}}$ shows noticeable deviation from a linear behavior. However, $v_{c, \mathrm{Ru}}$ can be plotted as a function of $N_B$, as it is done in Fig.~\ref{RLfig:combined_vc_Ru_B_as_sc_B}. In these coordinates both $v_{c, \mathrm{B}}$ and $v_{c, \mathrm{Ru}}$ change linearly with the amount of B on the surface. The slopes of both lines are also rather close, which suggests that the relative changes of characteristic velocities of Ru and B happen together, i.e. these changes depend on the surface composition but are not element-specific. It means that neutralization mechanism responsible for the matrix effect is of non-local nature, and also justifies the use of the Hagstrum model for the calculation of the inversed velocity from the incident ion energy~\cite{Zameshin2018a} in Fig.~\ref{RLfig:inv_velocity_plots}. 

Due to the abovementioned issues the work function cannot explain the observed matrix effect. It means that among two matrix effects observed for La in Chapter~\cite{Zameshin2018a}
the low work function effect (matrix effect in C-RN) is not involved; and oscillatory matrix effect matrix effect (matrix effect in qRCT) is not applicable in the current state as is expressed in changes of structuring of the ion yield curve, while our present matrix effect changes characteristic velocity. While this disqualifies matrix effect in qRCT, we should note that we cannot exclude presence of weak oscillations in Ru and B ion yields simply because of the lack of density of experimental points. However, none of these elements have atomic levels in proximity of He 1s level~\cite{Haynes2015}, which are required for presence of qRCT in the first place. 

\section{(Quasi-)resonant neutralization from the valence band} \label{RLsec:VB-qRN}

\begin{figure}[t]
    \centering
    \includegraphics[width=0.7\linewidth]{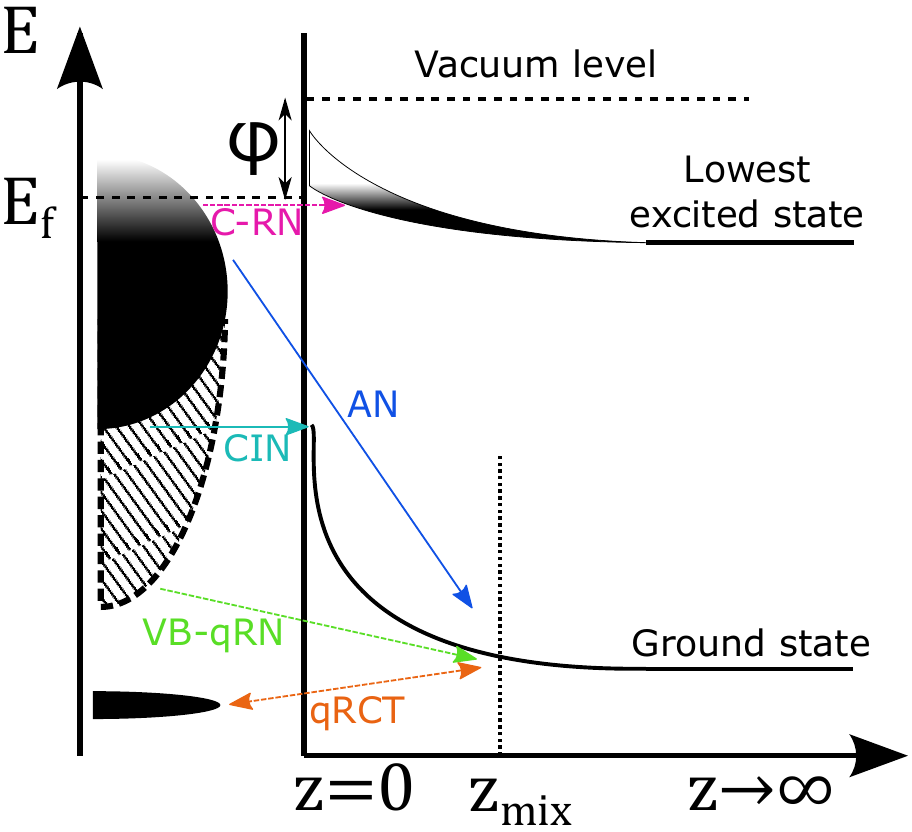}
    \caption{A schematic representation of possible neutralization mechanisms that can occur in LEIS. This drawing is based on the representation of Cortenraad~\cite{Cortenraad2002} with the addition of quasiresonant neutralization from the valence band of the target to the ground level of the projectile (VB-qRN).}
    \label{RLfig:LEIS_neutralization_mechanisms}
\end{figure}

Matrix effects in C-RN and qRCT are not the only two matrix effects recognized in LEIS field. There is also a separately classified matrix effect in graphitic carbon, which appears due to a neutralization mechanism solely mentioned in connection to this matrix effect: (quasi-)resonant neutralization from the valence band of the target to the ground state of the projectile~\cite{Brongersma1994,Prusa2015}  (Fig.~\ref{RLfig:LEIS_neutralization_mechanisms}). Due to the lack of standardized abbreviation we will call it VB-qRN in this thesis. VB-qRN is a non-local mechanism, because much like C-RN or CIN it originates from a wide band which spans multiple atoms in the solid. VB-qRN can only occur when the valence band is wide enough to become resonant with the ground state of the projectile; as demonstrated in~\cite{Prusa2015} this condition is satisfied for graphitic carbon. For most other materials, resonance between the bottom of the valence band and the projectile ground state is only possible at very small distances due to promotion of projectile levels in a close collision (CIN, collision-induced neutralization, or resonant neutralization in close collision). VB-qRN can be viewed as an intermediate case between CIN and qRCT. It is an extreme case of CIN with a threshold energy close to zero, when no promotion is required. It is also an extreme case of qRCT with a very wide band~\cite{Tsuneyuki1987}, which severely reduces the probability of reionization due to small hole lifetime within the band, and therefore makes for more efficient neutralization. It also causes severe dampening of ion yield oscillations as a function of inversed ion velocity~\cite{Prusa2015}.

VB-qRN relies on the unusually wide valence band of the target. We performed a literature search for calculations of the density of states (DOS) of valence electrons of different states of C, B and Ru, and the results are shown in Table~\ref{RLtab:bottom_valence_band}. Instead of comparing the full DOS, we only focus on the difference between the energy of the lowest state of the valence band and the Fermi level. We will call this energy $E_{\mathrm{lowest VB}}$. The uncertainty of such calculations is significant, therefore comparisons of $E_{\mathrm{lowest VB}}$ obtained with different techniques can be misleading. For an example of graphite, we were able to find $E_{\mathrm{lowest VB}}$ from 26 eV~\cite{Khvostov1996}, obtained from AES measurements, to 20 eV~\cite{Robertson1987}, calculated by the tight binding model. Such uncertainties cannot be avoided, but in several cases a relative comparison of $E_{\mathrm{lowest VB}}$ can be done for the calculations performed in the same source. 

Assuming typical value of work function of 4 eV, the valence band will be resonant with He 1s level as long as $E_{\mathrm{lowest VB}} > 24.6 - 4 = 20.6$ eV. By this measure graphitic C will be subject to VB-qRN mechanism. Metal carbides, e.g. MoC~\cite{Zaoui2005}, have much narrower valence band and will not have this neutralization channel. Therefore C in graphite will have much higher characteristic velocity than C in MoC, leading to a matrix effect.

\begin{table}[t]
\centering
\caption{Distance from the bottom of valence bands of several materials to the Fermi level ($E_{\mathrm{lowest VB}}$). The uncertainty of $E_{\mathrm{lowest VB}}$ from the literature is 0.5 eV, but much larger uncertainty comes from the calculations themselves. For elemental boron calculations show narrow sub-bands directly below valence band with bandgap $<2$ eV. It is unclear if these sub-bands become part of the valence band in amorphous B due to level broadening, therefore the positions of bottom of these sub-bands are shown in the table as well.}
\label{RLtab:bottom_valence_band}
\resizebox{0.8\textwidth}{!}{%
\begin{tabular}{|l|l|l|l|}
\hline
Material &
\begin{tabular}[c]{@{}l@{}}Bottom of valence band \\ $E_{\mathrm{lowest VB}}$, eV\end{tabular} &
\begin{tabular}[c]{@{}l@{}}If present: bottom \\ of sub-bands, eV\end{tabular} &
Source \\ \hline \hline
C, graphite & 26 & - & ~\cite{Khvostov1996} \\ \hline
C, graphite & 20 & - & ~\cite{Robertson1987} \\ \hline
C, diamond & 24 & - & ~\cite{Robertson1987} \\ \hline
C, amorph. 100\% sp2 & 21 & - & ~\cite{Robertson1987} \\ \hline
C, amorph. 0\% sp2 & 23.5 & - & ~\cite{Robertson1987} \\ \hline \hline
MoC & 11 & - & ~\cite{Zaoui2005} \\ \hline \hline
$\beta$-B & 16 & 19 & ~\cite{Prasad2005}, ~\cite{Imai2002} \\ \hline
$\alpha$-B & 22 & 26 & ~\cite{Armstrong1984} \\ \hline
a-B similar to $\alpha$-B & 16 & - & ~\cite{Durandurdu2017} \\ \hline
RuB\subt{4} & 20 & - & ~\cite{Wang2013} \\ \hline
RuB\subt{3} & 16 & - & ~\cite{Wang2013} \\ \hline
RuB\subt{2} & 14 & - & ~\cite{Wang2013} \\ \hline
Ru & 7.5 & - & ~\cite{Moreno-Armenta2007} \\ \hline

\end{tabular}%
}
\end{table}

Comparison of graphitic C with diamond and amorphous C~\cite{Robertson1987} shows that all of these states of C have $E_{\mathrm{lowest VB}}>20$ eV, which means that VB-qRN should be present in all elemental C and not only in graphite. We believe that all of the conclusions from the work of Pr{\r{u}}sa et al.~\cite{Prusa2015} can be applied to diamond and amorphous C as well. This allows to explain the behavior of Ru-C LEIS signal pair in Fig.~\ref{RLfig:matrix_effect_check}. The range of compositions starts with a pure amorphous C film, which has strong valence band neutralization, which reduces all LEIS signal (similar to situation with graphitic C). This effect is non-local, since it originates from the valence band, therefore Ru signal is affected as well. Ru carbides do not form at normal conditions~\cite{Harikrishnan2015full, Meschel2001}, therefore with increasing amount of Ru on the surface of the sample amount of electrons originating from valence band of C film proportionally reduces, and additional neutralization channel weakens. The decrease in the strength of neutralization causes an increase of the signals of both Ru and C, which is reason for a positive slope of the left side of Ru-C curve in Fig.~\ref{RLfig:matrix_effect_check}.

The right part of the curve behaves differently. Our assumption is that after a certain critical amount of Ru ($\vartheta_{\mathrm{Ru}} \gtrapprox 0.5$, achieved at $\thicksim 1$ nm of Ru) is deposited on the surface of the sample, no carbon atoms are bonded to each other in sufficient quantity to form a wide valence band, and additional neutralization channel disappears; matrix effect no longer present. This is represented by linear Ru-C dependence on the right side of Ru-C curve in Fig.~\ref{RLfig:matrix_effect_check}.

Table~\ref{RLtab:bottom_valence_band} shows that while the $E_{\mathrm{lowest VB}}$ of elemental B also varies between allotropes and different calculations, it can be expected in the range of 16 to 22 eV. Most of the calculations feature one or several sub-bands below the valence band, which may or may not be separated in the actual sample due to additional broadening of the levels. It means that bottom of the valence band of elemental B can potentially be in the strict resonance with He 1s level, and otherwise is in the quasiresonance with an energy mismatch of up to 4 eV. For qRN such energy mismatch still yields strong resonances~\cite{Brongersma2007, Zameshin2018a}, therefore we can expect VB-qRN in elemental B as well. B-rich ruthenium borides can have high $E_{\mathrm{lowest VB}}$~\cite{Wang2013}, which is noticeably decreasing with decreasing amount of B (from 20 for RuB\subt{4} to 14 for RuB\subt{2}). The strength of quasiresonance decreases with increasing energy mismatch, therefore VB-qRN will gets weaker with decreasing amount of B. Ru has a much smaller $E_{\mathrm{lowest VB}}$~\cite{Moreno-Armenta2007}, for which VB-qRN is impossible. It means that in the transition B - RuB\subt{x} - Ru strength of VB-qRN is constantly decreasing, leading to a constant decrease of characteristic velocity, leading to a matrix effect in Ru-B material combination.

With the help of VB-qRN we can explain additional details of the matrix effect in Ru-B that were observed in the section~\ref{RLsec:varE_measurements}. First of all, matrix effect is observed in both Ru and B LEIS signals, and relative change of characteristic velocity with composition is the same for B and Ru (Fig.~\ref{RLfig:combined_vc_Ru_B_as_sc_B}). VB-qRN is a non-local neutralization mechanism, which explains why it has the same effect on He ions scattered from both Ru and B atoms. Second, we can explain why $v_{c, \mathrm{B}}$ in Fig.~\ref{RLfig:vc_vs_comp_B} changes linearly with $N_\mathrm{B}$, while $v_{c, \mathrm{Ru}}$ deviates from a straight line in Fig.~\ref{RLfig:vc_vs_comp_Ru} but behaves more linearly in Fig.~\ref{RLfig:combined_vc_Ru_B_as_sc_B}. The reason for this is that VB-qRN originates from the bottom part of the valence band, which is formed by B-B and Ru-B bonds~\cite{Wang2013} and therefore depends on the amount of B more than Ru. 

In this section we established that the matrix effect in Ru-B mixtures occurs in a wide range of compositions (Fig.~\ref{RLfig:combined_vc_Ru_B_as_sc_B}), from approximately 1:1 mixture to pure B, but from this figure we are unable to say at which critical concentration of B in the mixture the matrix effect disappears. It appears that there are $v_{c, \mathrm{Ru}}$ and $v_{c, \mathrm{B}}$ change linearly with B concentration, but if the linear trend is extrapolated to $N_{\mathrm{B}} = 0$, it would lead to conclusion that a small amount of B added to Ru leads to change in $v_c$. However, at $E_{\mathrm{lowest VB}}$ or pure Ru a small change $E_{\mathrm{lowest VB}}$ should not lead to appearance of VB-qRN. Therefore, there must be a critical concentration at which the linear trend is no longer sustained. The only estimation of critical concentration remains from Fig.~\ref{RLfig:matrix_effect_check}, therefore a question of quantification of Ru-rich mixtures with LEIS can be considered open until more measurements of characteristic velocities are done.

\section{Discussion} \label{RLsec:Discussion}

Ru-B and Ru-C material combinations are not the only cases where matrix effects related to a band structure can be observed. Technically, \textit{all} borides and carbides should exhibit such matrix effects if corresponding pure elements are chosen as reference samples. This is simply because VB-qRN is present in elemental B and C but absent in other elements, therefore $P^+$ of at least one of the elements will be different in the mixture. The matrix effects should also propagate to situations where both elements have VB-qRN, for example a boron-carbon mixture (\BfC), because valence band shape and therefore neutralization efficiency in \BfC is different from any elemental B or C.

We expect that in a lot of cases matrix effects can be avoided by a proper choice of reference samples. For example, when measuring surface composition of certain metal carbide, one can use a pure surface of that metal as a reference for its LEIS signal, but reliable reference of C should only be obtained from extrapolation of a linear part of the pairwise signal comparison (Fig~\ref{RLfig:matrix_effect_check}), and not from any form of elemental carbon. This approach will work for any metal-rich surface composition that does not yet exhibit VB-qRN. For very carbon-rich surfaces in the presence of VB-qRN, the neutralization efficiency will be changing as a function of surface composition, and therefore the only reliable quantification method will be to measure the LEIS signal at different incident ion energies and extrapolate to infinite energy. With regard to transition metal borides with very delocalized B 2s-2p states, similar to RuB\subt{x} (OsB\subt{x}~\cite{Wang2013}, CoB\subt{x}, RhB\subt{x} and IrB\subt{x}~\cite{Wang2015}, ZrB\subt{2} and HfB\subt{2}~\cite{Zhang2008, Pan2015}), the compositional range of a constant neutralization efficiency might be narrow (see Section~\ref{RLsec:varE_measurements}), therefore the quantification of surface composition of transition metal borides will most probably require measurements at different energies. 

A slightly different situation arises with regard to hexaborides, as well as metal oxides and nitrides. Alkaline earth and rare earth hexaborides have much more localized B 2s-2p states~\cite{Huang2015, Bai2009}, which do not belong to a wide continuous valence band, yet are located at energies of 16 eV below the Fermi level. Metal oxides have O 2s states with similar energies: 19 eV for RuO\subt{2} in different phases~\cite{Mehtougui2012}, 19 eV for TiO\subt{2} in different phases~\cite{Chen2017}, 17 eV for ZrO\subt{2}~\cite{Zhang2016}, 19 eV for MgO and 20 eV for Al\subt{2}O\subt{3}~\cite{Kowalczyk1977ox}. Metal nitrides have N 2s states at 17 eV for Cr-Ti nitrides~\cite{Levy1999}, 16-18 eV for different Ru nitrides~\cite{Moreno-Armenta2007}, 16-17 eV for different phases of MoN~\cite{Kanoun2007} or 14 eV for MoN~\cite{Zaoui2005}, 16 eV for CrN and TiN~\cite{Zaoui2005}, 18 eV for VN~\cite{Zaoui2005}.

These isolated bands from B 2s-2p, O 2s or N 2s states can also be in a (quasi-)resonance with He 1s. This is the charge exchange process commonly known as qRN or qRCT. In the paper~\cite{Zameshin2018a}
we treated the target levels associated with qRCT as atomic levels, without taking their broadening into account. However, being a part of a solid, these levels can be treated as electron bands of a certain width. It is known that width of a band involved in a resonance affects the effectiveness of neutralization channel~\cite{Tsuneyuki1987}, which is expressed in dampening the oscillations of ion yield and increasing the characteristic velocity. It is also known that the radial distribution function of electron probability distribution is more narrow for d-electrons than for p- and s-electrons, which also means that energy splitting of d-electrons is lower and d-bands are more localized in space and in energy compared to p-bands. All strong oscillations in LEIS ion yield were produced by resonances with d-bands only~\cite{Brongersma2007}. In our earlier work~\cite{Zameshin2018a} we observed much weaker oscillation of ion yield in \Heplus scattering from La in different chemical states. We attributed the oscillations to quasiresonance between He 1s and La 5p levels, but the reason for severe dampening of oscillations was unclear. Now we can say that the broadening of p-orbitals compared to d-orbitals is the most likely explanation, that does not involve the use of an orbital symmetry rule, contrary to, for example, Souda et. al.~\cite{Souda1989}.

There is a lot of similarity between qRN from deep lying valence band states and qRN from relatively wide (i.e. s- or p-) non-valence bands. Changes in width and position of those bands can also affect the neutralization efficiency. Damplening of oscillations and increase of neutralization efficiency are caused by the same effect~\cite{Tsuneyuki1987}. In the earlier paper~\cite{Zameshin2018a}
we explained the changes in $v_{\mathrm{c}}$ of La by low work function matrix effect, however part of these changes might originate from qRN from B 2s-2p in \LaB and O 2s in \LaO. For the surfaces that are not subject to low work function matrix effect, for example for transition metal oxides, the change of characteristic velocity due to qRN can become dominant. Therefore, we can potentially expect qRN-related matrix effects in a variety of metal borides, carbides, oxides and nitrides. 

\begin{figure}
    \centering
    \begin{subfigure}[b]{0.475\textwidth}
        \centering
        \includegraphics[width=\textwidth]{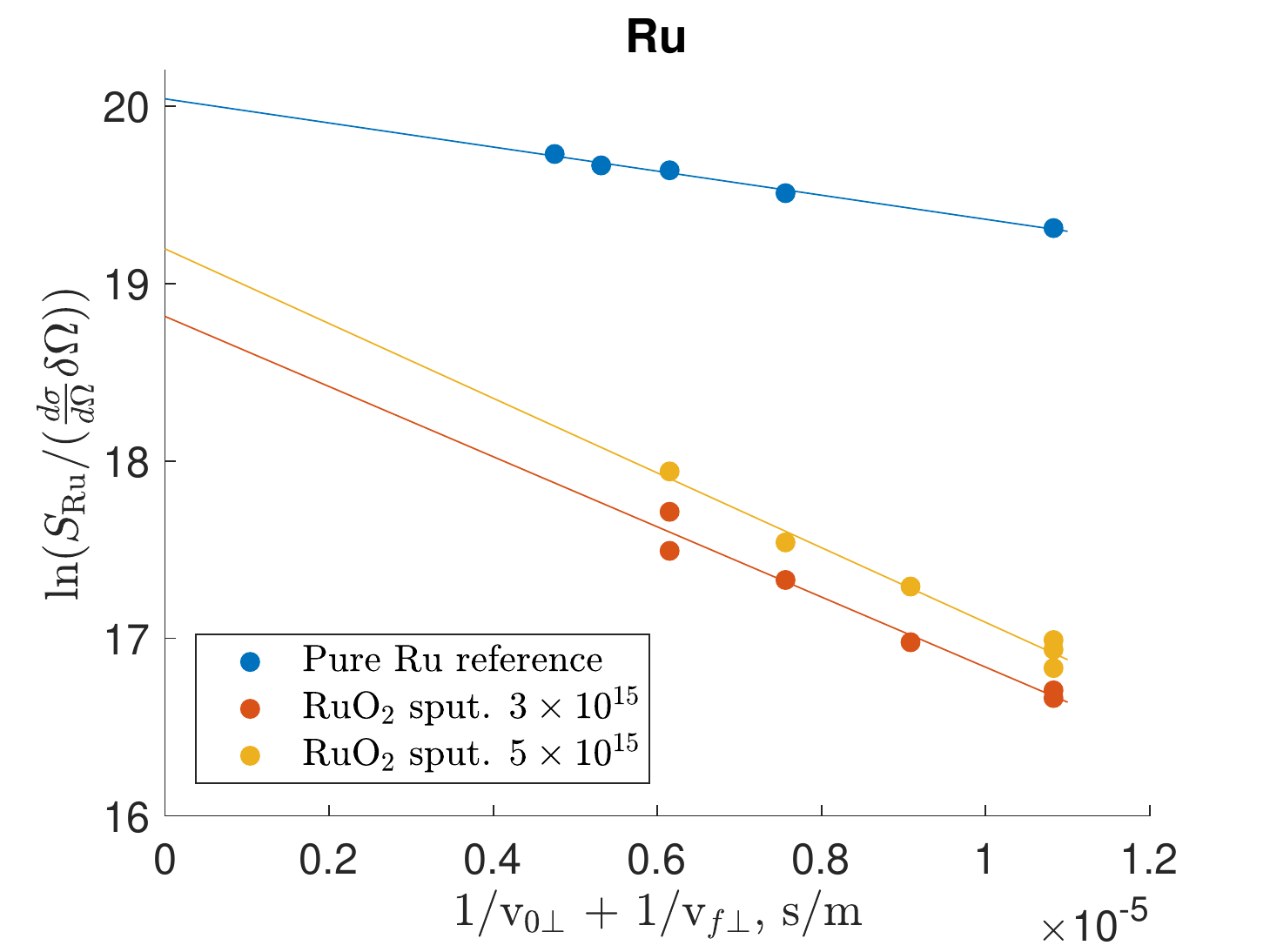}
        \phantomcaption    
        \label{RLfig:inv_velocity_plots_RuO_Ru}
    \end{subfigure}
    \hfill
    \begin{subfigure}[b]{0.475\textwidth}  
        \centering 
        \includegraphics[width=\textwidth]{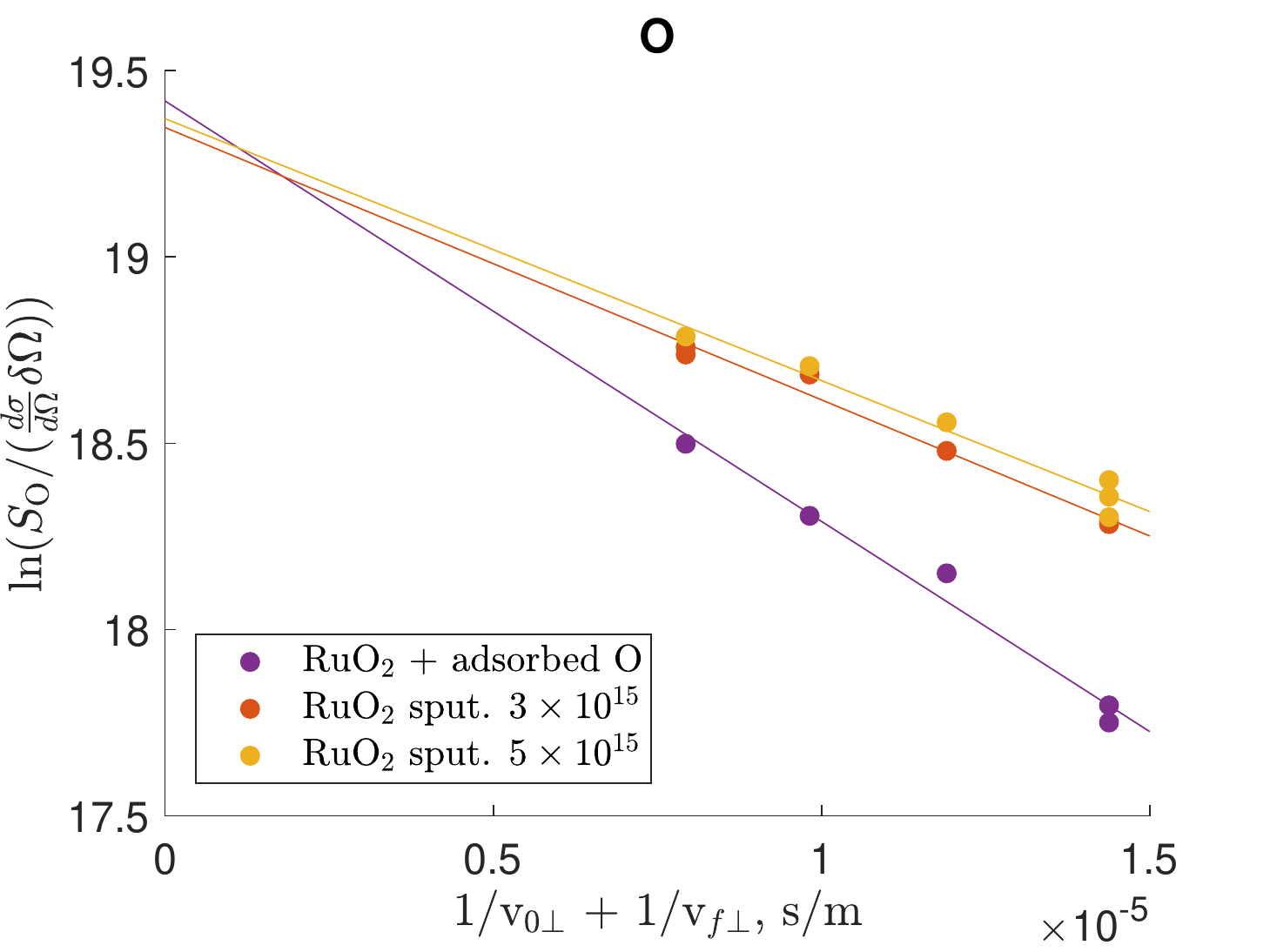}
        \phantomcaption   
        \label{RLfig:inv_velocity_plots_RuO_O}
    \end{subfigure}
    \caption
    {a) Dependence of logarithmic normalized LEIS signal of Ru on the inversed velocity of \Heplus ions, shown for pure and oxidized Ru. b) Dependence of logarithmic normalized LEIS signal of O on the inversed velocity of \Heplus ions, shown for different binding states of O: in RuO\subt{2} only and in RuO\subt{2} together with adsorbed O.} 
    \label{RLfig:inv_velocity_plots_RuO}
\end{figure}

To experimentally verify this prediction at least for Ru oxide, a simple proof-of-principle experiment was performed. A pure thin film of Ru was deposited by magnetron sputtering, and without breaking the vacuum exposed to atomic oxygen from a plasma source at a pressure of $1 \times 10^{-4}$ mbar for >1 hr at room temperature. The procedure should result in the formation of stoichiometric RuO\subt{2} thin film of $\approx$1 nm thickness. LEIS measurements of sample surfaces after the treatment did not show any quantifiable amount of Ru on the surface, which can be attributed to additional adsorbed O atoms on the surface. After mild 0.5 keV Ar\supt{+} sputtering LEIS signals of both Ru and O atoms can be reliably measured. The results of the measurements are shown in Fig.~\ref{RLfig:inv_velocity_plots_RuO} for both Ru and O peaks. There is a noticeable difference in characteristic velocities of pure Ru and oxidized Ru. There is also a difference between characteristic velocities of O from untreated surface (RuO\subt{2} + adsorbed O) and sputter-cleaned surface (RuO\subt{2} only). Understanding of the actual behavior of neutralization mechanisms in Ru oxides goes beyond the scope of this work, but this proof-of-principle experiment supports our prediction about existence of matrix effects related to neutralization O 2s levels.

\section{Conclusions} \label{RLsec:Conclusions}

In this work 3 keV \Heplus scattering was used to measure the surface composition of thin films of Ru on B, C and \BfC films at different stages of growth. Surface segregation of C on Ru and strong intermixing were observed in all cases. Matrix effects in \Heplus scattering from surfaces with a low Ru concentration ($\vartheta_{\mathrm{Ru}} \lessapprox 0.5$, achieved at $\thicksim 1$ nm of Ru) were observed. Dedicated studies of matrix effects were performed on the Ru-B material combination. After measuring the Ru and B signals with several \Heplus energies we observed a roughly linear increase of the characteristic velocities of Ru and B with the increase of B content. This phenomenon cannot be explained by a low work function matrix effect, which originates in from resonant neutralization from the conduction band of the target to excited levels of \Heplus. Yet, we explain this effect by the presence of quasiresonant neutralization from the valence band of B and RuB\subt{x} targets. Elemental B and Ru borides have wide valence bands with low lying states ($E_{\mathrm{lowest VB}} = 20$ to 14 eV) that can be in quasiresonance with the He 1s level, while the valence band of elemental Ru cannot participate in such quasiresonance ($E_{\mathrm{lowest VB}} = 7.5$ eV). This difference results in the changes of neutralization efficiency and therefore in LEIS signals, which gives the matrix effect in Ru-B surfaces. A similar matrix effect is observed for Ru-C surfaces. 

We further predict that qRN-related matrix effects of a similar mechanism can be applied to a much larger variety of compounds, i.e. metal borides, carbides, oxides and nitridies. A proof-of-principle experiment on oxidized Ru film shows a matrix effect in the Ru-O material combination as well, which can be attributed to qRN from O 2s levels. Further validation of this prediction necessitates systematic comparisons of characteristic velocities between pure elements and their compounds. We suggest that such measurements are necessary for reliable surface quantification in LEIS. We hope that this work will encourage more research on matrix effects originating from VB-qRN and qRN, and henceworth the limits of quantification of compounds by LEIS imposed by the presence of matrix effects.

\section{Acknowledgements} \label{RLsec:Acknowledgements}

This work was carried out in the Industrial Focus Group XUV Optics at the MESA+ Institute for Nanotechnology at the University of Twente, and we acknowledge the support of the industrial partners ASML, Carl Zeiss SMT, Malvern Panalytical, and TNO as well as the Province of Overijssel and the Dutch Organization for Scientific Research NWO. The authors express gratitude to Prof. Hidde Brongersma for helpful discussions as well as to Dr. Maria Berdova for providing the CVD B thin film sample.


\bibliographystyle{unsrtnat}

\end{document}